\documentclass{jpp}
\usepackage{graphicx}
\usepackage{epstopdf, epsfig,amsmath}

\renewcommand{\vec}[1]{\boldsymbol{#1}}

\newcommand{\diffp}[2]{\frac{\partial #1}{\partial #2}}
\shorttitle{ALPS: The Arbitrary Linear Plasma Solver}
\shortauthor{D.~Verscharen at al.}

\title{ALPS: The Arbitrary Linear Plasma Solver}

\author{D.~Verscharen\aff{1,2}
  \corresp{\email{d.verscharen@ucl.ac.uk}},
K.~G.~Klein\aff{3,4},
B.~D.~G.~Chandran\aff{2,5},
M.~L.~Stevens\aff{6},
C.~S.~Salem\aff{7},
\and S.~D.~Bale\aff{7,8}}

\affiliation{
\aff{1}Mullard Space Science Laboratory, University College London, Dorking RH5 6NT, UK
\aff{1}Space Science Center, University of New Hampshire, Durham, NH 03824, USA
\aff{3}Department of Climate and Space Sciences and Engineering, University of Michigan, Ann Arbor, MI 48109, USA
\aff{4}Lunar and Planetary Laboratory, University of Arizona, Tucson, AZ 85719, USA
\aff{5}Department of Physics, University of New Hampshire, Durham, NH 03824, USA
\aff{6}Harvard Smithsonian Center for Astrophysics, Cambridge, MA 02138, USA
\aff{7}Space Sciences Laboratory, University of California, Berkeley, CA 94720, USA
\aff{8}Department of Physics, University of California, Berkeley, CA 94720, USA}

\begin{document}

\maketitle
\begin{abstract}
The Arbitrary Linear Plasma Solver (ALPS) is a parallelised numerical code that solves the dispersion relation in a hot (even relativistic) magnetised plasma with an arbitrary number of particle species with arbitrary gyrotropic equilibrium distribution functions for any direction of wave propagation with respect to the background field. ALPS reads the background momentum distributions as tables of values on a $(p_{\perp},p_{\parallel})$ grid, where $p_{\perp}$ and $p_{\parallel }$ are the momentum coordinates in the directions perpendicular and parallel to the background magnetic field, respectively. We present the mathematical and numerical approach used by ALPS and introduce our algorithms for the handling of poles and the analytic continuation for the Landau contour integral.
We then show test calculations of dispersion relations for a selection of stable and unstable configurations in Maxwellian, bi-Maxwellian, $\kappa$-distributed, and J\"uttner-distributed plasmas. These tests demonstrate that ALPS derives reliable plasma dispersion relations. ALPS will make it possible to determine the properties of waves and instabilities in the non-equilibrium plasmas that are frequently found in space, laboratory experiments, and numerical simulations.
\end{abstract}

\section{Introduction}

The vast majority of the visible matter in the universe is in the plasma state. The solar wind is an example of such an astrophysical plasma. Due to its accessibility to spacecraft, it is the perfect environment for making comparisons between theoretical plasma-physics predictions and in-situ observations in the astrophysical context with access to wide scale separations  \citep[see, for example,][]{Marsch:2006}. 
Plasma can deviate from thermodynamic equilibrium if the relaxation due to 
 particle collisions occurs on timescales that are larger than the 
characteristic timescales of the collective plasma behaviour. Such a \emph{collisionless plasma} is characterised by non-Maxwellian features in its velocity distribution functions.
In the fast solar wind, this condition is frequently fulfilled, and, consequently, the
observed distribution functions often deviate from the entropically favoured Maxwellian shape
\citep{Vasyliunas:1968,Gosling:1981,Lui:1981,Marsch:1982,Marsch:1982a,Armstrong:1983,Lui:1983,Christon:1988,Williams:1988}. 
In particular, beams and temperature anisotropies are some of the observed features in the distributions of ions and electrons in the solar wind
\citep{Pilipp:1987a,Pilipp:1987b,Hellinger:2006,Marsch:2006,Bale:2009}. If these deviations from equilibrium are suitably extreme, the plasma becomes unstable and generates waves or non-propagating structures that react back upon the plasma to reduce the deviations from equilibrium 
\citep{Eviatar:1970,Schwartz:1980,Gary:1993,Hellinger:2011,Hellinger:2013}. 

The behaviour of plasma waves and instabilities is typically studied with the help of numerical codes that solve the hot-plasma dispersion relation. Traditionally, these codes (like WHAMP, PLUME, or NHDS) use a shifted bi-Maxwellian background distribution function as the zeroth-order description for the plasma state \citep{Roennmark:1982,Quataert:1998,Klein:2012,Verscharen:2013b}. For nearly collisionless plasmas, however, 
the bi-Maxwellian distribution function is a mathematical
convenience rather than a reliable representation of the 
true plasma distribution function, 
and many space-plasma observations show that the bi-Maxwellian representation is not accurate \citep{Hundhausen:1970,Leubner:1978,Marsch:1982,Pilipp:1987a,Marsch:2001,Stverak:2009}. 
Some previous approaches in non-Maxwellian solvers treated certain limits or geometries
\citep{dum80,Summers:1991,Summers:1994,Xue:1993,Xue:1996,Hellberg:2005, Cattaert:2007,Lazar:2009,Mace:2010,Lazar:2011, Galvao:2012,Xie:2013,Lazar:2014,gaelzer16,gaelzer16a} or faced challenges in the weakly-damped limit \citep{Hellinger:2011}. 

We present our numerical code ALPS (Arbitrary Linear Plasma Solver), which solves the full hot-plasma dispersion relation in a plasma consisting of an arbitrary number of particle species with arbitrary background distribution functions $f_{0j}$ and with arbitrary directions of wave propagation with respect to the uniform background magnetic field. ALPS is also able to solve the dispersion relation for relativistic plasmas.  \citet{matsuda92} developed a code similar to ALPS that calculates the dispersion relation in an arbitrary plasma with relativistic effects. Their code uses a cubic spline fit to both fill data gaps and approximate the analytic continuation, while ALPS uses a novel method called \emph{hybrid analytic continuation}. The spline method forfeits its accuracy for strongly damped solutions since the calculation of the dispersion relation requires the evaluation of the spline at a complex value that is distant from the real grid points by which the spline is supported. Our method does not suffer from this problem.
  \citet{astfalk17} also use a cubic-spline interpolation for the analytic continuation and as the basis for the integration in their code LEOPARD. This procedure allows for algebraic simplifications that enhance the speed of the integration significantly. LEOPARD, however, does not capture relativistic effects.

In Section~\ref{sect_DR}, we review the underlying theory of the hot-plasma dispersion relation. Section~\ref{sect_numeric} presents ALPS's numerical approach. In Section~\ref{sect_examples}, we compare ALPS results to known limits of the hot-plasma dispersion relation such as Maxwellian, bi-Maxwellian, $\kappa$-distributed, and relativistic pair plasmas. In Section~\ref{sect_conclusions}, we discuss our results and the applicability of ALPS to measured plasma distributions. The Appendix describes how ALPS solutions depend on the resolution of the background distributions, discusses of the Levenberg-Marquardt-fit routine used in our hybrid-analytic-continuation method, and describes our strategy for numerically refining coarse-grained distribution functions obtained from spacecraft measurements.

\section{The Linear Dispersion Relation of a Hot Plasma}\label{sect_DR}

In this section, we discuss the mathematical basis for the calculation of the hot-plasma dispersion relation following the presentation and notation of \citet{Stix:1992}. 
The determination of the kinetic wave dispersion relation in a hot plasma is based on the linearised set of Maxwell's 
equations and the linearised Vlasov equation \citep{Stix:1992,Gary:1993}.
A wave or instability is then associated with a first-order perturbation 
$\delta f_j$ in the distribution function of species $j$ about a prescribed 
time-averaged background distribution function $f_{0j}$,
\begin{equation}\label{linearization}
f_j(\vec x,\vec p, t)=f_{0j}(\vec p)+\delta f_j(\vec x,\vec p, t),
\end{equation} 
where $\vec x$ is the spatial coordinate and $\vec p$ is the momentum coordinate.
As with the distribution function $f_j$ in Equation~(\ref{linearization}), we take the magnetic field $\vec B$ to be the sum of a uniform background magnetic field $\vec B_0$ and a fluctuating magnetic field $\delta \vec B$. We assume that $\vec E=\delta \vec E$; i.e., the average electric field is zero. Linear theory expresses $\delta f_j$ as a function of $f_{0j}$ and the electromagnetic field components.  

The distribution function $f_j$ in a collisionless plasma evolves according to the Vlasov equation,
\begin{equation}\label{vlasov}
\diffp{f_j}{t}+\vec v\cdot \diffp{f_j}{\vec r}+q_j\left(\vec E+\frac{\vec v}{c}\times \vec B\right)\cdot\diffp{f_j}{\vec p}=0,
\end{equation}
where $q_j$ is the charge of a particle of species $j$, $c$ is the speed of light, and $\vec v$ is the velocity coordinate.  We assume that all fluctuating quantities behave like plane waves; i.e., $\propto \exp\left(i\vec k\cdot \vec x-i\omega t\right)$, where $\vec k$ is the wave vector and $\omega$ is the (complex) frequency. Linearising Equation~(\ref{vlasov}), using Faraday's law, and applying the method of characteristics, we obtain
\begin{multline}\label{deltafjone}
\delta f_j=-q_j e^{i\vec k\cdot \vec r-i\omega t}\int\limits_0^{\infty}\mathrm d\tau\, e^{i\alpha}\left\{\vphantom{\frac{A}{A}}E_xU\cos(\phi+\Omega_j\tau)+E_yU\sin(\phi+\Omega_j\tau)\right.\\
\left.+E_z\left[\diffp{f_{0j}}{p_{\parallel}}-V\cos(\phi-\vartheta+\Omega_j\tau)\right]\right\},
\end{multline}
where $\vec E=(E_x,E_y,E_z)$ is the electric field, $\phi$ is the azimuthal angle of the momentum vector $\vec p$, $\vartheta$ is the azimuthal angle of the wavevector $\vec k$, and the index $\perp$ ($\parallel$) refers to the direction perpendicular (parallel) with respect to the background magnetic field $\vec B_0$, 
\begin{equation}\label{omega_rel}
\Omega_j\equiv \frac{q_jB_0}{m_jc\sqrt{1+\left(p_{\perp}^2+p_{\parallel}^2\right)/m_j^2c^2}}
\end{equation}
is the relativistic gyrofrequency, $m_j$ is the rest mass of a particle of species $j$,
\begin{equation}
\alpha\equiv -\frac{k_{\perp}v_{\perp}}{\Omega_j}\left[\sin\left(\phi-\vartheta+\Omega_j\tau\right)-\sin \left(\phi-\vartheta\right)\right]+ \left(\omega-k_{\parallel}v_{\parallel}\right)\tau,
\end{equation}
\begin{equation}
U\equiv \diffp{f_{0j}}{p_{\perp}}+\frac{k_{\parallel}}{\omega}\left(v_{\perp}\diffp{f_{0j}}{p_{\parallel}}-v_{\parallel}\diffp{f_{0j}}{p_{\perp}}\right),
\end{equation}
and
\begin{equation}
V\equiv \frac{k_{\perp}}{\omega}\left(v_{\perp}\diffp{f_{0j}}{p_{\parallel}}-v_{\parallel}\diffp{f_{0j}}{p_{\perp}}\right).
\end{equation}
The first velocity moments of the distribution functions of all species define the current density $\vec j$ through
\begin{equation}
\vec j=\sum\limits_jq_j\int\mathrm d^3 \vec p\,\vec v \,\delta f_j=-\frac{i\omega}{4\pi}\sum\limits_j\vec \chi_j\cdot \vec E,
\end{equation}
where $\vec \chi_j$ is the contribution of species $j$ to the plasma susceptibility. Without loss of generality, we choose a cylindrical coordinate system in which $k_y=\vartheta=0$ and apply a set of Bessel-function identities in order to facilitate the integration over $\phi$ and $\tau$ in Equation~(\ref{deltafjone}). This allows us to rewrite the plasma susceptibilities  as  (provided that $\mathrm{Im}(\omega)>0$)
\begin{multline}\label{chis}
\vec {\chi}_j=\frac{\omega_{\mathrm pj}^2}{\omega\Omega_{0j}}\int\limits_0^{\infty}2\pi p_{\perp}\,\mathrm dp_{\perp}\int\limits_{-\infty}^{+\infty}\mathrm dp_{\parallel}\left[\hat{\vec e}_{\parallel}\hat{\vec e}_{\parallel}\frac{\Omega_j}{\omega}\left(\frac{1}{p_{\parallel}}\diffp{f_{0j}}{p_{\parallel}}-\frac{1}{p_{\perp}}\diffp{f_{0j}}{p_{\perp}}\right)p_{\parallel}^2\right.\\
\left.+\sum\limits_{n=-\infty}^{+\infty}\frac{\Omega_jp_{\perp}U}{\omega-k_{\parallel}v_{\parallel}-n\Omega_j}\mathsfbi T_n\right],
\end{multline}
where $\omega_{\mathrm pj}\equiv\sqrt{4\pi n_jq_j^2/m_j}$ is the plasma frequency of species $j$, $\Omega_{0j}\equiv q_jB_0/m_jc$ is the non-relativistic gyrofrequency, $n_j$ is the density of species $j$,  and the tensor $\mathsfbi T_n$ is defined as
\begin{equation}
\mathsfbi T_n\equiv \begin{pmatrix} \displaystyle \frac{n^2J_n^2}{z^2} & \displaystyle  \frac{inJ_nJ_n^{\prime}}{z} & \displaystyle \frac{nJ_n^2p_{\parallel}}{zp_{\perp}}\\[15pt]
\displaystyle -\frac{inJ_nJ_n^{\prime}}{z} & \displaystyle \left(J_n^{\prime}\right)^2 & \displaystyle -\frac{iJ_nJ_n^{\prime}p_{\parallel}}{p_{\perp}}\\[15pt]
\displaystyle \frac{nJ_n^2p_{\parallel}}{zp_{\perp}} & \displaystyle \frac{iJ_nJ_n^{\prime}p_{\parallel}}{p_{\perp}} & \displaystyle\frac{J_n^2p_{\parallel}^2}{p_{\perp}^2}
\end{pmatrix},
\end{equation}
where $z\equiv k_{\perp}v_{\perp}/\Omega_j$, and  $J_n\equiv J_n(z)$ is the $n$th-order Bessel function. For $\mathrm{Im}(\omega)\leq 0$, the integral over $p_{\parallel}$ is executed as the Landau integral after analytic continuation \citep[for details, see Chapt.~8 of][]{Stix:1992}.  Equation~(\ref{chis}) describes the susceptibility for a general background distribution function $f_{0j}$ in a relativistic plasma. The only assumptions are gyrotropy in $f_{0j}$ and small amplitudes in the fluctuations so that linearisation is applicable, and a uniform, stationary equilibrium. The numerical challenge in the solution of the plasma dispersion relation results from the integrals over $p_{\perp}$ and $p_{\parallel}$ in Equation~(\ref{chis}). We note that, in numerous classical codes for calculation of the linear hot-plasma dispersion relation \citep{Roennmark:1982,Gary:1993,Verscharen:2013a,Klein:2015a}, these integrals are greatly simplified by assuming that $f_{0j}$ is a (bi)-Maxwellian.

The dielectric tensor $\vec{\varepsilon}$ of the plasma is related to the plasma susceptibilities from Equation~(\ref{chis}) through
\begin{equation}
\vec{\varepsilon}=\mathsfbi 1+\sum\limits_j\vec{\chi}_j.
\end{equation}
Finally, combining Faraday's law and Amp\`ere's law leads to the wave equation,
\begin{equation}\label{dispersion}
\vec n\times\left(\vec n\times \vec E\right)+\vec{\varepsilon} \cdot \vec E\equiv \mathcal D \cdot \vec E=0,
\end{equation}
where $\vec n\equiv \vec kc/\omega$ is the index . By setting $\mathrm{det}\,\mathcal D=0$, we obtain the dispersion relations $\omega=\omega(\vec k)$ for non-trivial solutions to Equation~(\ref{dispersion}). We write these solutions in the form $\omega=\omega_{\mathrm r}+i\gamma$, where $\omega_{\mathrm r}=\mathrm{Re}(\omega)$ and $\gamma=\mathrm{Im}(\omega)$.

\section{Numerical Approach}\label{sect_numeric}


In order to find the solutions to the hot-plasma dispersion relation, ALPS determines the values of $\omega_{\mathrm r}$ and $\gamma$ that solve Equation~(\ref{dispersion}) for specified background distributions $f_{0j}$ at a given set of values for $\vec k$, $m_j$, $q_j$, $n_j$, and $v_{\mathrm A}/c$, where $v_{\mathrm A}\equiv B_0/\sqrt{4\pi n_{\mathrm p}m_{\mathrm p}}$. ALPS uses an efficient iterative Newton-secant algorithm to solve Equation~(\ref{dispersion}) based on an initial guess for $\omega_{\mathrm r}$ and $\gamma$ \citep{Press:1992}. The numerically challenging part for this calculation is the evaluation of $\vec{\chi}_j$ in Eqs.~(\ref{chis}). In the following, we present ALPS's strategy for this evaluation in the non-relativistic case. We discuss the extension to relativistic cases with poles in the integration domain in Section~\ref{resonances}, which is equivalent to the non-relativistic case with the exception that the coordinate system is transformed from $(p_{\perp},p_{\parallel})$ to $(\Gamma,\bar p_{\parallel})$ and that Equation~(\ref{chisrel}) below is used instead of Equation~(\ref{chis}).

We prescribe the shape of $f_{0j}$ in input files for each species (called ``$f_{0}$ table'') as an ASCII table that  lists $p_{\perp}$, $p_{\parallel}$, and the associated values of $f_{0j}$. From this table, we calculate $\partial f_{0j}/\partial p_{\perp}$ and $\partial f_{0j}/\partial p_{\parallel}$ on the same grid as the $f_0$ table using second-order finite differencing. The resolution of the $f_0$ table is given by $n_{\perp}$ points in the $p_{\perp}$-direction and $n_{\parallel}$ points in the $p_{\parallel}$-direction. The table spans from $p_{\perp}=0$ to $p_{\perp}=P_{\max,\perp j}$ in the perpendicular direction and from $p_{\parallel}=-P_{\max,\parallel j}$ to $p_{\parallel}=P_{\max,\parallel j}$ in the parallel direction.

The integration in Equation~(\ref{chis}) allows us to integrate separately and independently for each $n$ and $j$. This provides us with a very natural way to parallelise the calculation scheme by assigning the separate integrations to different processors. We use MPI for the parallelisation. The integrating nodes return their contributions to $\vec{\chi}_j$ to the master node, which then sums up the contributions, determines the value of $\vec{\varepsilon}$, and updates the values of $\omega_{\mathrm r}$ and $\gamma$ through a Newton-secant step. The updated values for $\omega_{\mathrm r}$ and $\gamma$ are then returned to the integrating nodes, which afterwards evaluate the integration of their updated contribution to $\vec{\chi}_j$. We evaluate all values of $n$ up to a value of $\pm n_{\max}$, which is determined as the value of $n$ for which the maximum value of $|J_{n}|$ is smaller than the user-defined parameter $J_{\max}$. The necessary value of $n_{\max}$ depends on the wavenumber, the direction of propagation of the treated wave, and the thermal speeds of the plasma components. In bi-Maxwellian codes under typical solar-wind conditions, the accuracy of the dispersion relation is better than $\Delta |\omega|/|\omega|\sim 10^{-5}$ for $J_{\max}\sim 10^{-45}$ (which typically corresponds to  $n_{\max}\gtrsim 10$ at proton scales).

We use a standard two-dimensional trapezoidal integration scheme to integrate over $p_{\perp}$ and $p_{\parallel}$. However, this scheme breaks down near the poles of the integrand in Equation~(\ref{chis}) and requires a special treatment of the analytic continuation when $\gamma\leq 0$. In the remainder of this section, we discuss our strategies to resolve these numerical difficulties.

\subsection{Integrating Near Poles}\label{resonance_int}

A challenge concerning the numerical integration is the treatment of the poles that occur in the term proportional to $\mathsfbi T_n$ in Equation~(\ref{chis}). 
The integrals in question are of the form
\begin{equation}\label{Iint}
I(p_{\perp})=\int\limits_{-\infty}^{+\infty}\mathrm dp_{\parallel}\frac{\Omega_jU\mathsfbi T_n}{\omega-k_{\parallel}v_{\parallel}-n\Omega_j}\equiv \int\limits_{-\infty}^{+\infty}\mathrm dp_{\parallel} G(p_{\perp},p_{\parallel})
\end{equation}
for $\gamma>0$.
For sufficiently small $\gamma$, the denominator in Equation~(\ref{Iint}) can become very small along the real $p_{\parallel}$ axis so that the grid sampling leads to large numerical errors in the integration.  To describe how we evaluate these integrals, we first rewrite the integral in Equation~(\ref{Iint}) in the more generic form
\begin{equation}
\mathcal I=\int\limits_{-\infty}^{+\infty}\mathrm dx\frac{g(x)}{x-t_{\mathrm r}-it_{\mathrm i}},
\end{equation}
where $x$, $t_{\mathrm r}$, and $t_{\mathrm i}$ are real, $g(x)$ is a smooth function, and the integration is performed along the real axis. 
We choose a symmetric interval $[t_{\mathrm r}-\Delta,t_{\mathrm r}+\Delta]$ around $t_{\mathrm r}$ where $\Delta \ll g(t_{\mathrm r})/g^{\prime}(t_{\mathrm r})$, and write
\begin{equation}
\mathcal I=\int\limits_{t_{\mathrm r}-\Delta}^{t_{\mathrm r}-\Delta}\mathrm dx\frac{g(x)}{x-t_{\mathrm r}-it_{\mathrm i}}+\text{rest},
\end{equation}
where ``rest'' refers to the integration outside the interval $[t_{\mathrm r}-\Delta,t_{\mathrm r}+\Delta]$. We define a function $f(x)$ to be odd with respect to $t_{\mathrm r}$ if $f(x)=-f(2t_{\mathrm r}-x)$, and even with respect to $t_{\mathrm r}$ if $f(x)=f(2t_{\mathrm r}-x)$. Following \citet{Longman:1958} and \citet{Davis:1984}, we then separate the integrand into its odd and even parts with respect to $t_{\mathrm r}$ as
\begin{multline}\label{iintlong}
\mathcal I=\frac{1}{2}\int\limits_{t_{\mathrm r}-\Delta}^{t_{\mathrm r}+\Delta}\mathrm dx\left[\frac{g(x)}{x-t_{\mathrm r}-it_{\mathrm i}}-\frac{g(2t_{\mathrm r}-x)}{-x+t_{\mathrm r}-it_{\mathrm i}}\right] \\
 +\frac{1}{2}\int\limits_{t_{\mathrm r}-\Delta}^{t_{\mathrm r}+\Delta}\mathrm dx\left[\frac{g(x)}{x-t_{\mathrm r}-it_{\mathrm i}}+\frac{g(2t_{\mathrm r}-x)}{-x+t_{\mathrm r}-it_{\mathrm i}}\right]  +\text{rest}.
\end{multline}
The integrand in the first integral in Equation~(\ref{iintlong}) is odd with respect to $t_{\mathrm r}$ and thus vanishes after the integration over the symmetric interval around $t_{\mathrm r}$. The second integral, on the other hand, is even with respect to $t_{\mathrm r}$ and thus
\begin{equation}\label{ifinal}
\mathcal I= \int\limits_{t_{\mathrm r}}^{t_{\mathrm r}+\Delta}\mathrm dx\left[\frac{g(x)}{x-t_{\mathrm r}-it_{\mathrm i}}-\frac{g(2t_{\mathrm r}-x)}{x-t_{\mathrm r}+it_{\mathrm i}}\right]  +\text{rest}.
\end{equation}
We define $\Delta$ through a user-defined parameter $n_{\mathrm I}$ so that $\Delta\equiv n_{\mathrm I}\,\Delta p_{\parallel}$. We then define $\delta\equiv \Delta/n_{\mathrm P}$, where $n_{\mathrm P}$ is another user-defined parameter. 
Except for cases in which $|t_{\mathrm i}|$ is extremely small, we apply a trapezoidal integration over $n_{\mathrm P}$ steps of width $\delta$ to the integral in Equation~(\ref{ifinal}). The smoothness of $g(x)$ allows us to expand $g(x)$ around the nearest grid point of the $n_{\mathrm I}$ grid points in the interval $[t_{\mathrm r},t_{\mathrm r}+\Delta]$ using a Taylor series. By taking $\Delta p_{\parallel}$ to be sufficiently small, we can retain just the first two terms in the series without losing significant accuracy. Since the integral in Equation~(\ref{ifinal}) does not converge numerically if $|t_{\mathrm i}|$ is extremely small, we implement the following procedure when $|t_{\mathrm i}|\leq t_{\mathrm{lim}}$, where $t_{\mathrm{lim}}$ is a user-defined parameter. We first rewrite Equation~(\ref{ifinal}) using truncated Taylor expansions of $g(x)$ and $g(2t_{\mathrm r}-x)$ around $x=t_{\mathrm r}$ as
\begin{equation}\label{ianalyt}
\mathcal I= \int\limits_{t_{\mathrm r}}^{t_{\mathrm r}+\Delta}\mathrm dx\left[\frac{2it_{\mathrm i}g(t_{\mathrm r})}{\left(x-t_{\mathrm r}\right)^2+t_{\mathrm i}^2}+\frac{2g^{\prime}(t_{\mathrm r})\left(x-t_{\mathrm r}\right)^2}{\left(x-t_{\mathrm r}\right)^2+t_{\mathrm i}^2}\right]  +\text{rest}.
\end{equation}
We determine $g(t_{\mathrm r})$ and $g^{\prime}(t_{\mathrm r})$ through linear interpolation between the neighbouring grid points to $t_{\mathrm r}$. The term proportional to $g^{\prime}(t_{\mathrm r})$ in Equation~(\ref{ianalyt}) converges numerically for any value of $t_{\mathrm i}$. We set the term proportional to $g(t_{\mathrm r})$ equal to its small-$t_{\mathrm i}$ limit, namely 
\begin{equation}
\int\limits_{t_{\mathrm r}}^{t_{\mathrm r}+\Delta}\mathrm dx  \frac{2it_{\mathrm i}g(t_{\mathrm r})}{\left(x-t_{\mathrm r}\right)^2+t_{\mathrm i}^2} =  i\pi g(t_{\mathrm r})\mathrm{sgn}(t_{\mathrm i}).
\end{equation}
We use this method for both the integration of $\vec{\chi}_j$ near poles and the principal-value integration that is necessary if $\gamma=0$.

\subsection{Analytic Continuation}\label{hac}

If $\gamma\leq 0$, the integration in Equation~(\ref{chis}) requires an analytic continuation into the complex plane. If $f_{0j}$ were given as a closed algebraic expression, the analytic continuation would simply entail the evaluation of $f_{0j}(p_{\perp},p_{\parallel})$ at a complex value for $p_{\parallel}$ in the non-relativistic case. In our case, however, $f_{0j}$ is only defined on a real grid in $p_{\perp}$ and $p_{\parallel}$, yet the analytic continuation of $f_{0j}$ is still uniquely defined. This leads to the known mathematical problem of \emph{numerical analytic continuation} \citep{Cannon:1965,Reichel:1986,Fujiwara:2007,Fu:2012,Zhang:2013,Kranich:2014}. Our solution for this problem is our \emph{hybrid analytic continuation} scheme. We note that this approach is only relevant for damped modes, i.e., $\gamma\leq 0$. 

Landau's rule of integration around singularities \citep{Landau:1946,Lifshitz:1981} leads to the following three cases with the appropriate residues for the evaluation of $I(p_{\perp})$ for general $\gamma$:
\begin{equation}\label{landau}
I(p_{\perp})=\int_{C_{\mathrm L}}\mathrm dp_{\parallel}G(p_{\perp},p_{\parallel})=\begin{cases}
\int\limits_{-\infty}^{+\infty}\mathrm dp_{\parallel}G(p_{\perp},p_{\parallel}) & \text{if } \gamma>0,\\
\mathcal P \int\limits_{-\infty}^{+\infty}\mathrm dp_{\parallel}G(p_{\perp},p_{\parallel}) + i\pi \sum \mathrm{Res}_{A}(G)& \text{if } \gamma=0,\\
\int\limits_{-\infty}^{+\infty}\mathrm dp_{\parallel}G(p_{\perp},p_{\parallel}) + 2 i\pi \sum \mathrm{Res}_{A}(G) & \text{if } \gamma<0,\\
\end{cases}
\end{equation}
where $C_{\mathrm L}$ is the contour of the Landau integration, which lies below the complex poles in the integrand. The integrations on the right-hand side of Equation~(\ref{landau}) are performed along the real axis, and $\mathcal P$ indicates the principal-value integral. The sum sign indicates the summation over the residues of all poles $A$ of the function $G$. In a non-relativistic plasma, $G$ has one simple pole, and thus
\begin{equation}
\sum \mathrm{Res}_A(G) = -\frac{m_j}{\left |k_{\parallel}\right |} \left.\Omega_{j}U\mathsfbi T_n\right|_{p_{\parallel}=p_{\mathrm {pole}}},
\end{equation}
where $p_{\mathrm{pole}}=m_j\left(\omega-n\Omega_j\right)/k_{\parallel}$ is the parallel momentum associated with pole $A$.

It is a common approach to decompose the background distribution functions in terms of analytical expressions and then to evaluate these at the complex poles. Complete orthogonal basis functions such as Hermite, Legendre, or Chebyshev polynomials are the prime candidates for such a decomposition since they can represent $f_{0j}$ to an arbitrary degree of accuracy \citep{robinson90,weideman95,Xie:2013}. These approaches are useful when $f_{0j}$ deviates only slightly from a Maxwellian. They require, however, very high orders of decomposition and are thus slow in the presence of typical structures that we see in the solar wind such as a proton core-beam configuration. Therefore, they are unsuitable for ALPS's purpose, and we pursue a different approach, which we call the \emph{hybrid analytic continuation}. The basic idea behind this approach is to integrate $I$ numerically along the real axis whenever possible and to resort to an algebraic function for the sole purpose of the evaluation of $\mathrm{Res}_A(G)$  when necessary. 

For the determination of an appropriate algebraic function, ALPS allows the user to choose an arbitrary combination of fit functions to represent $f_{0j}$ and automatically evaluates the fits before the integration begins. The code evaluates the fits separately at each $p_{\perp}$, so that no assumption is made as to the structure of $f_{0j}$ in the $p_{\perp}$-direction.  ALPS uses these functions only if a pole is within the integration domain and only if $\gamma\leq 0$. The intrinsic fit functions that the code can combine include a Maxwellian distribution,
\begin{equation}\label{maxwell}
f_{0j}=\frac{n_{j}}{\pi^{3/2}m_j^3w_{\perp j}^2w_{\parallel}}\exp\left(-\frac{p_{\perp}^2}{m_j^2w_{\perp j}^2}-\frac{\left(p_{\parallel}-m_jU_j\right)^2}{m_j^2w_{\parallel j}^2}\right),
\end{equation}
where  $w_{\perp j}\equiv \sqrt{2k_{\mathrm B}T_{\perp j}/m_j}$ ($w_{\parallel j}\equiv \sqrt{2k_{\mathrm B}T_{\parallel j}/m_j}$) is the  thermal speed of species $j$ in the direction perpendicular (parallel) with respect to $\vec B_0$, $T_{\perp j}$ ($T_{\parallel j}$) is the temperature of species $j$ perpendicular (parallel) to $\vec B_0$, $k_{\mathrm B}$ is the Boltzmann constant, and $U_j$ is the $\vec B_0$-parallel drift speed of species $j$; a $\kappa$-distribution \citep{Summers:1994,Astfalk:2015}, 
\begin{multline}
f_{0j}=\frac{n_{j}}{m_j^3w_{\perp j}^2w_{\parallel j}}\left[\frac{2}{\pi(2\kappa-3)}\right]^{3/2}\frac{\tilde{\Gamma}(\kappa+1)}{\tilde{\Gamma}(\kappa-1/2)}\\
\times \left\{1+\frac{2}{2\kappa-3}\left[\frac{p_{\perp}^2}{m_j^2w_{\perp j}^2}+\frac{(p_{\parallel}-m_jU_j)^2}{m_j^2w_{\parallel j}^2}\right]\right\}^{-(\kappa+1)};
\end{multline}
and a J\"uttner distribution \citep{juettner11,chacon10},
\begin{equation}\label{juett}
f_{0j}=\frac{n_j}{2\pi m_j^3c w_j^2 K_2\left(w_j^2/2c^2\right) }\exp\left(-2\frac{c^2}{w_j^2}\sqrt{1+\frac{|\vec p|^2}{m_j^2c^2}}\right);
\end{equation}
where $\kappa$ is the $\kappa$-index, $\tilde{\Gamma}$ is the gamma function, and $K_2$ is the modified Bessel function of the second kind. The J\"uttner distribution is the thermodynamic-equilibrium distribution if $k_{\mathrm B}T_{j}\gtrsim m_jc^2$. The exponential in Equation~(\ref{juett}) reduces to the Maxwellian $\exp(-v^2/w_j^2)$ with a different $\vec p$-independent normalisation factor for $p^2/m_j^2c^2\ll 1$. We use an automated Levenberg--Marquardt-fit algorithm \citep{Levenberg:1944,Marquardt:1963} and describe the details of the fit routine in Appendix \ref{sect_LMfit}.

\subsection{The Poles in a Relativistic Plasma}\label{resonances}

The analytic continuation and pole handling in the relativistic case entail a further complication due to the non-trivial $\vec p$-dependence of the resonant denominator in Equation~(\ref{chis}) \citep{buti62,lerche68}. We define a plasma to be relativistic when there is a significant number of particles at relativistic velocities. This can be the case in plasmas with relativistic temperatures ($k_{\mathrm B}T_j\gtrsim m_jc^2$) or in plasmas with relativistic beams ($P_{j}\gtrsim m_jc$, where $P_j$ is the drift momentum). Using the relativistic expression for $\Omega_j$ in Equation~(\ref{omega_rel}) shows that we can write for the pole of the function under the integral sign in Equation~(\ref{Iint}) 
\begin{equation}\label{reso1}
\frac{1}{\omega-k_{\parallel}v_{\parallel}-n\Omega_j}=-\frac{1}{k_{\parallel}}\frac{\Gamma m_j}{\displaystyle \left(p_{\parallel}-\frac{\omega}{k_{\parallel}}\Gamma m_j+n\frac{\Omega_{0j}}{k_{\parallel}}m_{j}\right)},
\end{equation}
where 
\begin{equation}\label{lorentz}
\Gamma\equiv \sqrt{1+\frac{p_{\perp}^2+p_{\parallel}^2}{m_j^2c^2}}
\end{equation}
 is the Lorentz factor. We define the dimensionless parallel momentum $\bar p_{\parallel}\equiv p_{\parallel}/m_{j}c$. The dimensionless parallel momentum associated with the relativistic pole is given by
\begin{equation}\label{resmom}
\bar p_{\mathrm{pole}} = \Gamma \frac{\omega}{k_{\parallel}c}-\frac{n\Omega_{0j}}{k_{\parallel}c}.
\end{equation}

We apply the technique proposed by \citet{Lerche:1967} to transform Equation~(\ref{chis}) from the $(p_{\perp},p_{\parallel})$ coordinate system to the $(\Gamma,\bar p_{\parallel})$ coordinate system \citep[see also][]{Swanson:2002,Lazar:2006,Lopez:2014,lopez16}. This transformation yields
\begin{multline}\label{chisrel}
\vec{\chi}_j=2 \pi m_j^3 c^3\frac{\omega_{\mathrm pj}^2}{\omega \Omega_{0j}}\int \limits _1^{\infty} \mathrm d\Gamma \int \limits_{-\sqrt{\Gamma^2-1}} ^{+\sqrt{\Gamma^2-1}}\mathrm d\bar p_{\parallel}\left[\hat{\vec e}_{\parallel} \hat{\vec e}_{\parallel} \frac{\Omega_{0j}}{\omega}\bar p_{\parallel} \diffp{f_{0j}}{\bar p_{\parallel}}\right. \\
\left.-\sum \limits_{n=-\infty}^{+\infty}\frac{\Omega_{0j}}{k_{\parallel }c}\left(\diffp{f_{0j}}{\Gamma}+\frac{k_{\parallel}c}{\omega}\diffp{f_{0j}}{\bar p_{\parallel}}\right)\frac{1}{\bar p_{\parallel}-\Gamma \frac{\omega}{k_{\parallel}c}+\frac{n\Omega_{0j}}{k_{\parallel}c}} \bar{\mathsfbi T}_n\right],
\end{multline}
where 
\begin{equation}
\bar{\mathsfbi T_n}\equiv \begin{pmatrix} \displaystyle \frac{n^2J_n^2}{\bar z^2} & \displaystyle  \frac{inJ_nJ_n^{\prime}}{\bar z} \bar p_{\perp} & \displaystyle \frac{nJ_n^2\bar p_{\parallel}}{\bar z}\\[15pt]
\displaystyle -\frac{inJ_nJ_n^{\prime}}{\bar z}\bar p_{\perp} & \displaystyle \left(J_n^{\prime}\right)^2 \bar p_{\perp}^2 & \displaystyle -iJ_nJ_n^{\prime}\bar p_{\parallel}\bar p_{\perp}\\[15pt]
\displaystyle \frac{nJ_n^2\bar p_{\parallel}}{\bar z} & \displaystyle iJ_nJ_n^{\prime}\bar p_{\parallel}\bar p_{\perp} & \displaystyle J_n^2\bar p_{\parallel}^2 
\end{pmatrix},
\end{equation}
$\bar p_{\perp}\equiv \sqrt{\Gamma^2-1-\bar p_{\parallel}^2}$, $\bar z\equiv k_{\perp}c/\Omega_{0j}$, and the Bessel functions are evaluated as $J_n\equiv J_n(\bar z\bar p_{\perp})$. Whenever ALPS performs a relativistic calculation and 
\begin{equation}
-P_{\max,\parallel j} \leq \mathrm{Re}\left(p_{\mathrm{pole}}\right)\leq +P_{\max,\parallel j},
\end{equation}
the code automatically transforms from $(p_{\perp},p_{\parallel})$ to $(\Gamma,\bar p_{\parallel})$ coordinates and applies the polyharmonic spline algorithm described in Appendix~\ref{interpolation} to create an equally spaced and homogeneous grid in $(\Gamma,\bar p_{\parallel})$ coordinates. In this coordinate system, we perform the integration near poles and the analytic continuation in the same way as described in Sects.~\ref{resonance_int} and \ref{hac}, but using the relativistic parallel momentum associated with the pole from Equation~(\ref{resmom}).
For reasons of numerical performance, we use the integration based on Equation~(\ref{chisrel}) only if there is a pole within the integration domain. Otherwise, we employ the faster integration method based on Equation~(\ref{chis}) even in the relativistic case.

\section{Test Cases and Results}\label{sect_examples}

In this section, we compare ALPS with known reference cases based on either our own or previously published results. 

\subsection{Maxwellian Distributions}

\begin{figure}
  \centerline{\includegraphics[trim={0.7cm 1cm 0.2cm 0.1cm},clip,width=\textwidth]{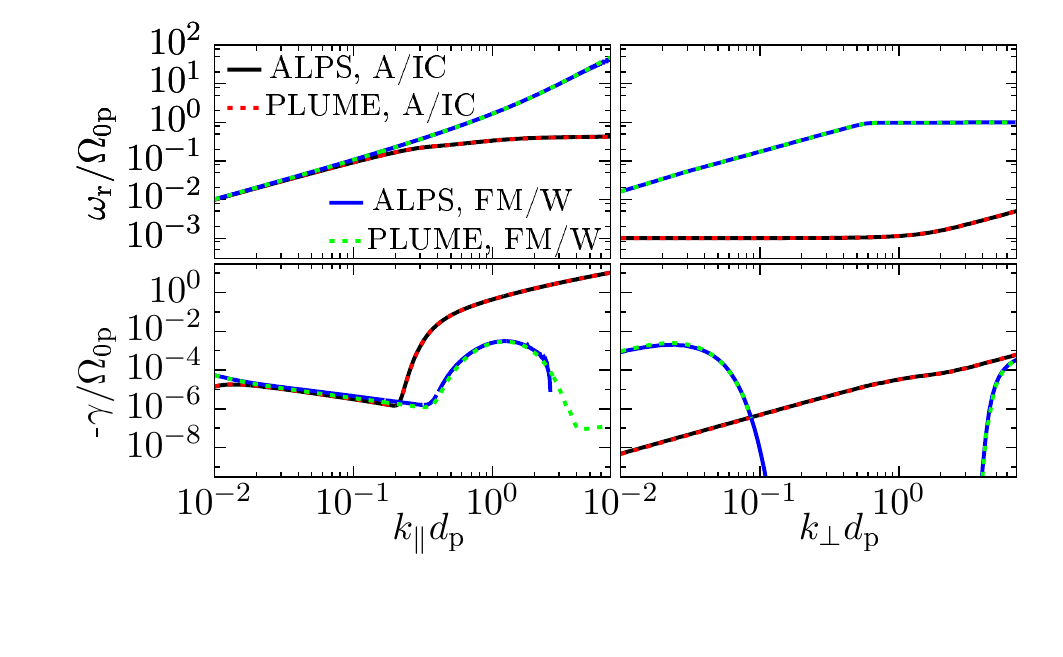}}
  \caption{Dispersion relations for the A/IC wave and the FM/W wave  in a Maxwellian plasma in quasi-parallel (left) and quasi-perpendicular (right) propagation. For the calculations shown on the left, we keep $k_{\perp}d_{\mathrm p}=10^{-3}$ constant and scan through $k_{\parallel}$. For the calculations shown on the right, we keep $k_{\parallel}d_{\mathrm p}=10^{-3}$ constant and scan through $k_{\perp}$. The A/IC mode in quasi-perpendicular propagation corresponds to the kinetic Alfv\'en wave (KAW) at $k_{\perp}d_{\mathrm p}\gtrsim 1/\sqrt{\beta_{\parallel \mathrm p}}$. We compare ALPS with the standard Maxwellian solutions from PLUME for an electron-proton plasma with the same plasma parameters. Both numerical models agree well in both the real part $\omega_{\mathrm r}$ of the frequency and its imaginary part $\gamma$.}
\label{fig_par_example}
\end{figure}

There are numerous codes for the hot-plasma dispersion relation in a plasma with Maxwellian or bi-Maxwellian background distributions. We use our code PLUME \citep{Klein:2015a} for an electron-proton plasma and calculate the dispersion relations of Alfv\'en/ion-cyclotron (A/IC) and fast-magnetosonic/whistler (FM/W) waves. We then set up Maxwellian $f_{0}$ tables with the same parameters as those used with PLUME and calculate the dispersion relations based on these $f_{0j}$ tables with ALPS. 
We compare the PLUME and ALPS results for quasi-parallel and quasi-perpendicular propagation in Figure~\ref{fig_par_example}. The panels show both the real part of the frequency $\omega_{\mathrm r}$ and its imaginary part $\gamma$ as functions of the parallel and perpendicular wavenumbers, respectively. We use $\beta_{\perp j}=\beta_{\parallel j}=1$ for both protons and electrons, and $v_{\mathrm A}/c=10^{-4}$, where $\beta_{\perp j}\equiv w_{\perp j}^2/v_{\mathrm A}^2$ and $\beta_{\parallel j}\equiv w_{\parallel j}^2/v_{\mathrm A}^2$. We normalise all frequencies in units of the proton cyclotron frequency $\Omega_{0\mathrm p}$ and all length scales in units of the proton skin depth $d_{\mathrm p}\equiv v_{\mathrm A}/\Omega_{0\mathrm p}$. The momentum-space resolution for the ALPS calculation in the quasi-parallel limit is $n_{\perp}=320$, $n_{\parallel}=640$,  $P_{\max,\parallel\mathrm p}=8m_{\mathrm p}v_{\mathrm A}$, and $P_{\max,\parallel\mathrm e}=0.19m_{\mathrm p}v_{\mathrm A}$. In the quasi-perpendicular limit, we use $n_{\perp}=240$, $n_{\parallel}=480$,  $P_{\max,\parallel\mathrm p}=6m_{\mathrm p}v_{\mathrm A}$, and $P_{\max,\parallel\mathrm e}=0.14m_{\mathrm p}v_{\mathrm A}$. In both cases, we set $P_{\max,\parallel j}=P_{\max,\perp j}$, $J_{\max}=10^{-45}$, $n_{\mathrm I}=5$,  $n_{\mathrm p}=100$, and $T_{\mathrm{lim}}=0.01$. We study the accuracy of the results depending on the resolution in Appendix~\ref{app_resolution_max}. Figure~\ref{fig_par_example} shows that ALPS reproduces these Maxwellian examples very well. We note that these plasma parameters represent typical solar-wind conditions at 1~au. 

\begin{figure}
  \centerline{\includegraphics[trim={0.3cm 0.2cm 0.5cm 0.1cm},clip,width=\textwidth]{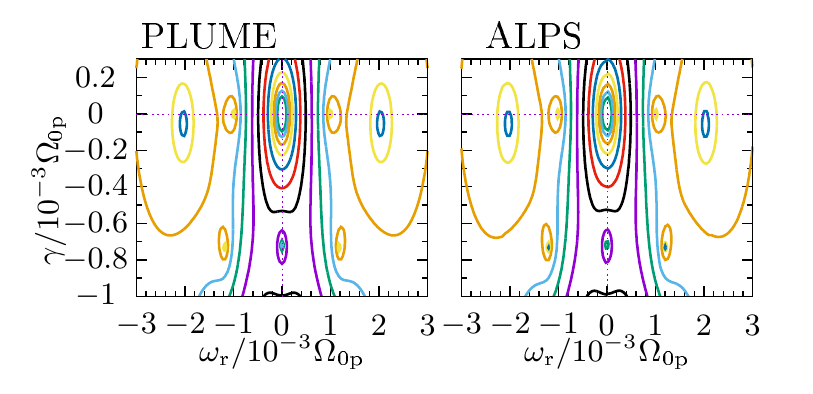}}
  \caption{Comparison of dispersion maps from PLUME (left) and ALPS (right) for $k_{\perp} d_{\mathrm p}=k_{\parallel}d_{\mathrm p}=10^{-3}$. The lines show isocontours of constant $\lg\left|\mathrm{det}\,\mathcal D\right|$. Minima in these maps correspond to solutions to the hot-plasma dispersion relation.}
\label{fig_maps}
\end{figure}

In order to illustrate another representation of the plasma dispersion relation, we show a comparison of \emph{dispersion maps} from PLUME and ALPS in Figure~\ref{fig_maps}. Dispersion maps are diagrams of isocontours of constant $\lg\left|\mathrm{det}\,\mathcal D\right|$, where $\mathcal D$ is the tensor from Equation~(\ref{dispersion}), in the $\omega_{\mathrm r}$--$\gamma$ plane. They are a useful tool to find the initial guesses for $\omega_{\mathrm r}$ and $\gamma$ for the Newton-secant root-finding search.  Although the calculation of a dispersion map still requires the calculation of all $\vec{\chi}_j$, it does not entail the application of the Newton-secant root-finding algorithm. Solutions to the hot-plasma dispersion relation appear as minima in these diagrams. We use a  Maxwellian plasma model with $\beta_{\perp j}=\beta_{\parallel j}=1$ for both protons and electrons, $k_{\perp} d_{\mathrm p}=k_{\parallel}d_{\mathrm p}=10^{-3}$, and $v_{\mathrm A}/c=10^{-4}$. For the ALPS calculation, we use $n_{\perp}=240$, $n_{\parallel}=480$, $P_{\max,\parallel\mathrm p}=P_{\max,\perp\mathrm p}=6m_{\mathrm p}v_{\mathrm A}$,  $P_{\max,\parallel\mathrm e}=P_{\max,\perp\mathrm e}=0.14m_{\mathrm p}v_{\mathrm A}$,  $J_{\max}=10^{-45}$, $n_{\mathrm I}=5$,  $n_{\mathrm p}=100$, and $T_{\mathrm{lim}}=0.01$. Both the ALPS and the PLUME calculations reveal seven solutions to the dispersion relation. We note that the point $\omega_{\mathrm r}=\gamma=0$ is a maximum and does not represent a solution to the dispersion relation. The solutions at $\omega_{\mathrm r}=\pm 10^{-3}\Omega_{0\mathrm p}$ and $\gamma=-2.3\times 10^{-10}\Omega_{0\mathrm p}$ are the forward and backward propagating A/IC waves. The solutions at $\omega_{\mathrm r}=\pm 2\times 10^{-3}\Omega_{0\mathrm p}$ and $\gamma=-5.4\times 10^{-5}\Omega_{0\mathrm p}$ are the forward and backward propagating FM/W waves. The solutions at $\omega_{\mathrm r}=\pm 1.2\times 10^{-3}\Omega_{0\mathrm p}$ and $\gamma=-7.3\times 10^{-4}\Omega_{0\mathrm p}$ are the forward and backward propagating slow waves (ion-acoustic waves). Lastly, the solution at $\omega_{\mathrm r}=0$ and $\gamma=-7.2\times 10^{-4}\Omega_{0\mathrm p}$ is the non-propagating slow mode, which is sometimes denoted `entropy mode', \citep{Verscharen:2016,verscharen17}. The comparison of both panels in Figure~\ref{fig_maps} shows that ALPS reproduces these seven plasma modes under typical solar-wind conditions in the Maxwellian limit.

\subsection{Anisotropic Bi-Maxwellian Distributions}

\begin{figure}
  \centerline{\includegraphics[trim={0.7cm 1.0cm 0.2cm 0.cm},clip,width=\textwidth]{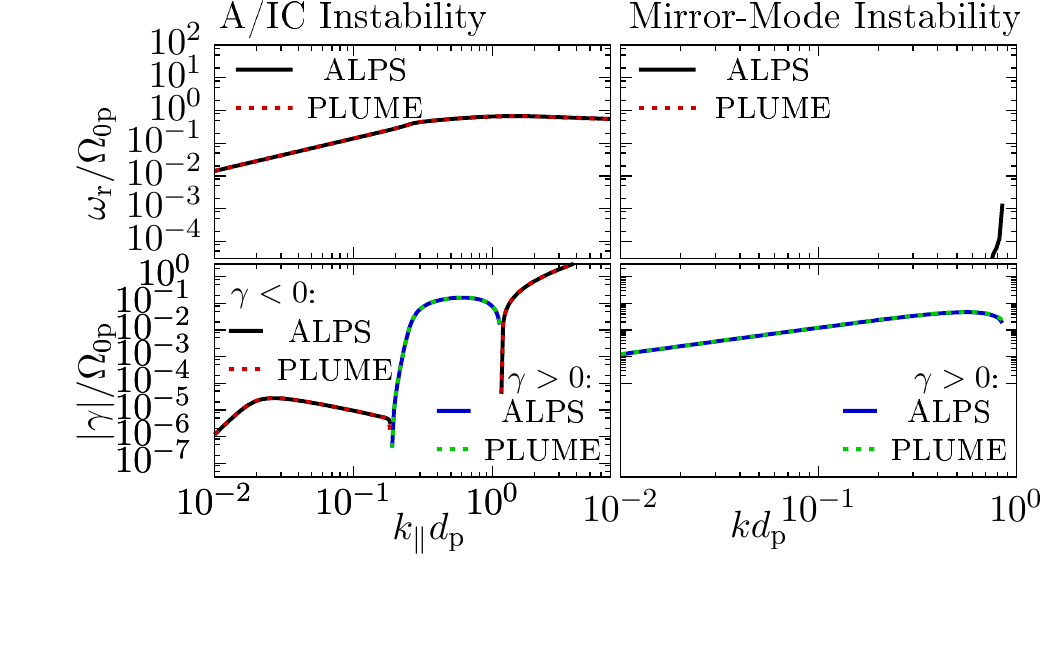}}
  \caption{Comparison of dispersion relations for the A/IC instability (left) and the mirror-mode instability (right) from PLUME and ALPS. We use $T_{\perp\mathrm p}/T_{\parallel\mathrm p}=3$. For the calculation of the A/IC instability, we keep $k_{\perp}d_{\mathrm p}=10^{-3}$ constant and scan through $k_{\parallel}$. For the calculation of the mirror-mode instability, we keep $\theta=75^{\circ}$ constant and scan through $|\vec k|$.}
\label{fig_instable_example}
\end{figure}

PLUME, like most other standard hot-plasma dispersion-relation solvers, also allows us to use anisotropic bi-Maxwellian representations for the background distribution functions. Such a configuration can lead to instability if the temperature anisotropy exceeds the threshold for an anisotropy-driven plasma instability. As an example for a propagating instability, we calculate the dispersion relation for the parallel A/IC instability \citep{Harris:1961,Davidson:1975,Yoon:2010}, and as an example for a non-propagating instability, we calculate the dispersion relation for the mirror-mode instability \citep{Rudakov:1961,Tajiri:1967,Southwood:1993}. The thresholds for both of these instabilities fulfil $T_{\perp\mathrm p}>T_{\parallel\mathrm p}$. 
For this demonstration, we use PLUME to calculate $\omega_{\mathrm r}$ and $\gamma$ as functions of the wavenumber in a plasma with bi-Maxwellian protons and Maxwellian electrons using $\beta_{\parallel\mathrm p}=\beta_{\parallel\mathrm e}=\beta_{\perp\mathrm e}=1$, $T_{\perp\mathrm p}/T_{\parallel\mathrm p}=3$, and $v_{\mathrm A}/c=10^{-4}$. We then set up bi-Maxwellian $f_{0}$ tables with the same parameters and calculate the dispersion relations for both instabilities with ALPS. We show the results in Figure~\ref{fig_instable_example}. For the ALPS calculation, we use $n_{\perp}=320$, $n_{\parallel}=640$, and $P_{\max,\parallel\mathrm p}=8m_{\mathrm p}v_{\mathrm A}$, $P_{\max,\perp\mathrm p}=13.9m_{\mathrm p}v_{\mathrm A}$,  $P_{\max,\parallel\mathrm e}=P_{\max,\perp\mathrm e}=0.19m_{\mathrm p}v_{\mathrm A}$, $J_{\max}=10^{-45}$, $n_{\mathrm I}=5$,  $n_{\mathrm p}=100$, and $T_{\mathrm{lim}}=0.01$. We study the accuracy of these results depending on the resolution in Appendix~\ref{app_resolution_instabilities}.

Both PLUME and ALPS show that the A/IC wave and the mirror mode are unstable in different wave-vector ranges for the given parameter set. The good agreement between the PLUME solutions and the ALPS solutions shows that ALPS successfully calculates the dispersion relations of both instabilities in a bi-Maxwellian plasma.

\subsection{Anisotropic $\kappa$-Distributions}

\begin{figure}
\centerline{\includegraphics[trim={0.8cm 1.cm 0.6cm 0.1cm},clip,width=0.7\textwidth]{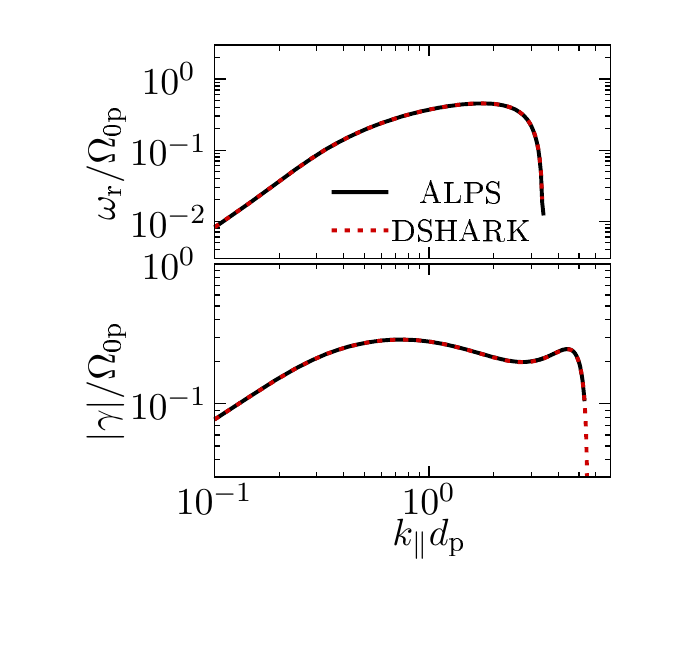}}
  \caption{Comparison of dispersion relations of the FM/W instability in a $\kappa$-distributed plasma from DSHARK and ALPS. In both plasma models, we keep $\theta=0.001^{\circ}$ constant and scan through $k_{\parallel}$. The top panel shows the real part of the wave frequency, and the bottom panel shows its imaginary part.}
\label{fig_kappa_example}
\end{figure}

\citet{Astfalk:2015} developed the code DSHARK to calculate dispersion relations in plasmas with bi-$\kappa$-distributions. As one example, these authors discuss the FM/W instability in an anisotropic electron-proton plasma with $\kappa_{\mathrm p}=\kappa_{\mathrm e}=8$, $\beta_{\parallel\mathrm p}=2$, $\beta_{\parallel\mathrm e}=4$, $T_{\perp\mathrm p}/T_{\parallel\mathrm p}=0.4$, and $T_{\perp\mathrm e}/T_{\parallel\mathrm e}=0.5$ \citep[see Figure 1 from][]{Astfalk:2015}. The angle between $\vec k$ and $\vec B_0$ is constant for this calculation and set to $\theta=0.001^{\circ}$. We use DSHARK to reproduce this test case and set up $\kappa$-distributed $f_0$ tables with the same parameters in order to compare the DSHARK results with ALPS. We show this comparison in Figure~\ref{fig_kappa_example}. In ALPS, we use $n_{\perp}=400$, $n_{\parallel}=800$, $P_{\max,\parallel\mathrm p}=10m_{\mathrm p}v_{\mathrm A}$, $P_{\max,\perp\mathrm p}=6.32m_{\mathrm p}v_{\mathrm A}$, $P_{\max,\parallel\mathrm e}=0.33m_{\mathrm p}v_{\mathrm A}$,  $P_{\max,\perp\mathrm e}=0.23m_{\mathrm p}v_{\mathrm A}$, $J_{\max}=10^{-45}$, $n_{\mathrm I}=5$, $n_{\mathrm p}=500$, $T_{\mathrm{lim}}=0.01$, and $v_{\mathrm A}/c=10^{-4}$.

ALPS reproduces the DSHARK results for the FM/W instability well. The results also agree with the previous work by \citet{Lazar:2011}.

\subsection{Relativistic J\"uttner Distributions}

\begin{figure}
\centerline{\includegraphics[trim={0.7cm 0.cm 0.3cm 0.cm},clip,width=\textwidth]{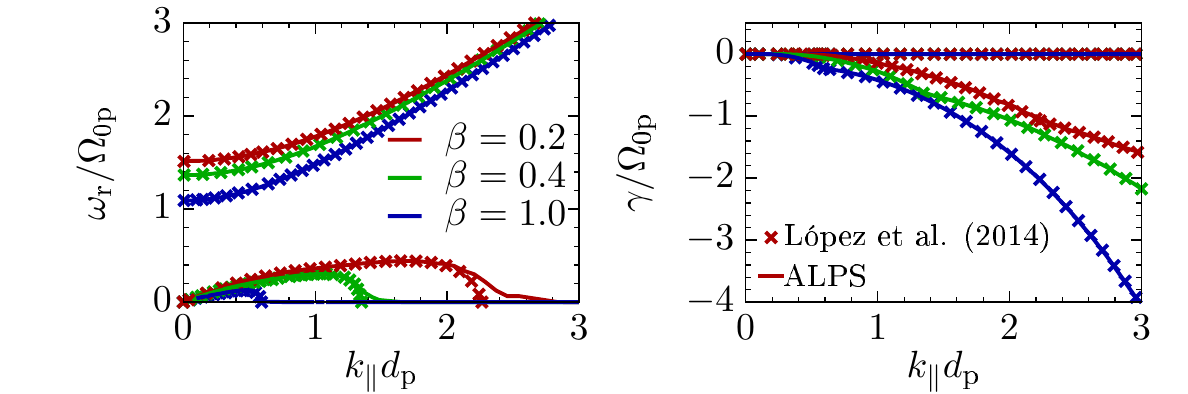}}
  \caption{Dispersion relations of the quasi-parallel A/IC wave (solutions at low $\omega_{\mathrm r}$) and the Ordinary wave (solutions at high $\omega_{\mathrm r}$) in a relativistic electron-positron pair plasma with J\"uttner distributions.  We keep $k_{\perp}d_{\mathrm p}=10^{-3}$ constant and scan through $k_{\parallel}$. The left panel shows the real part of the frequency, and the right panel shows the imaginary part of the frequency. The lines show ALPS solutions, and the crosses show the results from Figure~1 of \citet{Lopez:2014}. The three colours correspond to $\beta_{\mathrm p}=\beta_{\mathrm e}=0.2$ (red), $\beta_{\mathrm p}=\beta_{\mathrm e}=0.4$ (green), and $\beta_{\mathrm p}=\beta_{\mathrm e}=1.0$ (blue) in both panels and for both modes.}
\label{fig_relativistic}
\end{figure}

As one example for a dispersion relation in a relativistic plasma, we reproduce the results by \citet{Lopez:2014} for an electron-positron pair plasma with a J\"uttner distribution using $v_{\mathrm A}/c=1$, $m_{\mathrm p}=m_{\mathrm e}$, $\beta_{\parallel \mathrm p}=\beta_{\parallel \mathrm e}=(0.2,0.4,1.0)$ and $T_{\perp j}=T_{\parallel j}$ for both positrons and electrons. We set up a J\"uttner-distributed $f_0$ table with the same parameters and calculate the dispersion relations of the A/IC wave and the Ordinary wave (O-mode) in the plasma, keeping the perpendicular wavenumber constant at $k_{\perp}d_{\mathrm p}=10^{-3}$. We use $n_{\perp}=30$, $n_{\parallel}=60$,  $P_{\max}=5m_{\mathrm p}v_{\mathrm A}$,  $J_{\max}=10^{-45}$, $n_{\mathrm I}=5$,  $n_{\mathrm p}=300$, and $T_{\mathrm{lim}}=0.01$. Our interpolation method transforms the $(p_{\perp},p_{\parallel})$ grid to the $(\Gamma,\bar p_{\parallel})$ grid with $n_{\Gamma}=500$ and $n_{\bar p_{\parallel}}=500$ steps in $\Gamma$ and $\bar p_{\parallel}$, respectively. We show the results in Figure~\ref{fig_relativistic}.
\citet{Lopez:2014} show their results for these parameters in their Figure~1. Our comparison with the ALPS dispersion relation in Figure~\ref{fig_relativistic} shows a good agreement and confirms our relativistic model. The deviation between the results from \citet{Lopez:2014} and ALPS is only visible in the real part of the frequency at the large-$k_{\parallel}$/low-$\omega_{\mathrm r}$ end of the A/IC branches.

\section{Discussion and Conclusions}\label{sect_conclusions}

ALPS solves the relativistic and non-relativistic hot-plasma dispersion relations in a plasma with arbitrary background distribution functions. We have benchmarked ALPS against existing codes by comparing dispersion relations for waves and instabilities in Maxwellian, bi-Maxwellian, $\kappa$-distributed, and relativistic J\"uttner-distributed plasmas. In all cases, we find that ALPS agrees well with existing codes. This finding encourages us to apply ALPS to yet unexplored plasma environments in future work.

An important application of ALPS will be the analysis of distribution functions measured by spacecraft in the solar wind. ALPS includes the necessary numerical framework to preprocess and format the spacecraft data so that they can serve as  $f_0$ tables for direct input (see Appendix~\ref{interpolation}). Especially, the upcoming missions Solar Orbiter and Parker Solar Probe will deliver plasma measurements with unprecedented energy and time resolution in the solar wind that will serve as the ideal input for ALPS. The vast majority of previous kinetic studies of waves and instabilities relied on bi-Maxwellian fits to the observed distribution functions and the use of a standard bi-Maxwellian code to solve the hot-plasma dispersion relation such as WHAMP, NHDS, or PLUME. Our approach allows us, however, to relax the bi-Maxwellian assumption and to analyse the plasma behaviour more realistically.
Future comparisons of the results from standard codes such as PLUME with the results from ALPS will help to evaluate the quality of the previous bi-Maxwellian approaches and to refine our understanding of the role of instabilities in collisionless plasmas based on the actual distribution functions.  For instance, our knowledge of the realistic value of certain instability thresholds is still very limited. Some in-situ observations of kinetic plasma features in the solar wind lie above the thresholds of kinetic instabilities when calculated based on bi-Maxwellian background  distributions \citep[see, for example,][]{Isenberg:2012}. The general conjecture is, however, that the plasma is limited by the lowest instability threshold. A more realistic calculation based on the actual distribution functions may resolve this discrepancy. This concept applies, for example, to anisotropy-driven instabilities such as the A/IC instability \citep{Hellinger:2006,Bale:2009,Maruca:2012} or beam-driven instabilities such as the FM/W instability \citep{Reisenfeld:2001,Verscharen:2013,Verscharen:2013b}. Also non-thermal electron configurations, which are known to carry a significant heat flux into the solar wind, require a non-bi-Maxwellian representation for the determination of the relevant instabilities that limit their heat flux \citep{Feldman:1975,Pilipp:1987a,Pulupa:2011,Salem:2013}. Another field of application of ALPS is the study of highly non-thermal plasma configurations related to reconnection events \citep{Phan:2006,Gosling:2007,Gosling:2007a,Egedal:2012,Egedal:2013}. We also emphasise the applicability of ALPS for the determination of dispersion relations using distributions from numerical plasma simulations. Particle-in-cell or Eulerian plasma codes generate data directly suitable as $f_{0}$ tables for ALPS. Some of these numerical simulations use (realistically or artificially) relativistic plasma conditions. Therefore, ALPS's ability to include relativistic effects will be very useful for the study of the wave properties and the stability of simulated plasmas.

Our resolution studies in Appendix~\ref{app_resolution} offer some insight into the necessary resolution of the $f_0$ tables for a reliable determination of the plasma dispersion relation. In the shown applications, a minimum resolution of about $n_{\perp}=40$ and $n_{\parallel}=80$ has proven to be necessary for a good agreement between ALPS and the test results for (bi-)Maxwellian distributions. In a future extension of ALPS, we will include Nyquist's method to automatically determine the stability of directly observed distribution functions \citep{Klein:2017}.

\acknowledgements

We appreciate helpful comments and contributions from Sergei Markovskii and Thomas Brackett. The ALPS collaboration appreciates support from NASA grant NNX16AG81G. We present more details about the numerics on the website www.alps.space. The ALPS source code will be made publicly available on this website after our initial science phase. Computations were performed on Trillian, a Cray XE6m-200 supercomputer at UNH supported by the NSF MRI program under grant PHY-1229408. D.V. was supported by the STFC Ernest Rutherford Fellowship ST/P003826/1. B.D.G.C. was supported in part by NASA grants NNX15AI80G and NNX17AI18G and NSF grant PHY-1500041.

\appendix

\section{Resolution Studies}\label{app_resolution}

In order to understand the required resolution of the $f_{0}$ tables for calculations with ALPS, we compare results from PLUME with results from ALPS for the same plasma parameters using different resolutions in this appendix. For all calculations, we use $\beta_{\parallel\mathrm p}=\beta_{\parallel\mathrm e}=1$, $T_{\perp\mathrm e}/T_{\parallel\mathrm e}=1$,  $J_{\max}=10^{-45}$, $n_{\mathrm I}=5$,  $n_{\mathrm p}=100$, and $T_{\mathrm{lim}}=0.01$. We use $P_{\max,\parallel\mathrm p}$ as a free parameter and set $n_{\parallel}=2n_{\perp}$, $P_{\max,\parallel \mathrm e}=P_{\max,\perp \mathrm e}=P_{\max,\parallel \mathrm p}\sqrt{m_{\mathrm e}/m_{\mathrm p}}$, and $P_{\max,\perp\mathrm p}=P_{\max,\parallel\mathrm p}\sqrt{T_{\perp\mathrm p}/T_{\parallel\mathrm p}}$. 
We define the resolution in momentum as $\Delta w_{j}\equiv P_{\max,\parallel j}/(n_{\parallel}m_jw_{\parallel j})$ and the resolution in frequency as $\Delta \omega_{\mathrm r}/\omega_{\mathrm r}\equiv \left| \omega_{\mathrm r,\mathrm{ALPS}}-\omega_{\mathrm r,\mathrm{PLUME}}\right|/\omega_{\mathrm r,\mathrm{PLUME}}$, where $\omega_{\mathrm r,\mathrm{ALPS}}$ is the solution from ALPS, and $\omega_{\mathrm r,\mathrm{PLUME}}$ is the solution from PLUME.  For $\kappa$-distributions, the appropriate resolution depends on both $\beta_j$ and $\kappa$. Instead of giving general guidelines for the resolution, we, therefore, recommend case-by-case convergence studies when calculating dispersion relations in plasmas with $\kappa$-distributions.

\subsection{Maxwellian Distributions}\label{app_resolution_max}

\begin{figure}
  \centerline{\includegraphics[trim={0.4cm 0.5cm 0.2cm 0.1cm},clip,width=\textwidth]{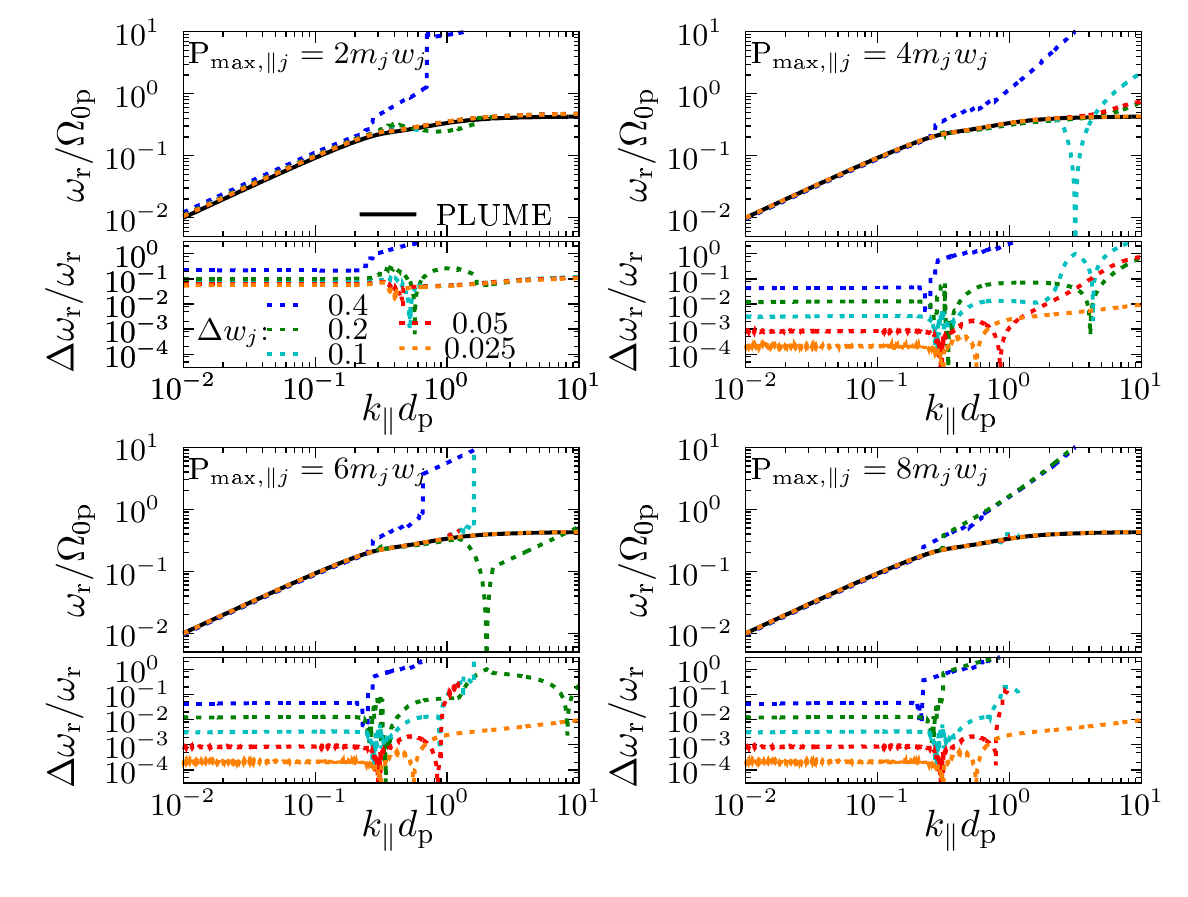}}
  \caption{Resolution study for the real part of the frequency for the A/IC-wave solution in quasi-parallel propagation. We keep $k_{\perp}d_{\mathrm p}=10^{-3}$ constant and scan through $k_{\parallel}$. }
\label{fig_ALPS_convergence_a_par_omega}
\end{figure}

\begin{figure}
  \centerline{\includegraphics[trim={0.4cm 0.5cm 0.2cm 0.1cm},clip,width=\textwidth]{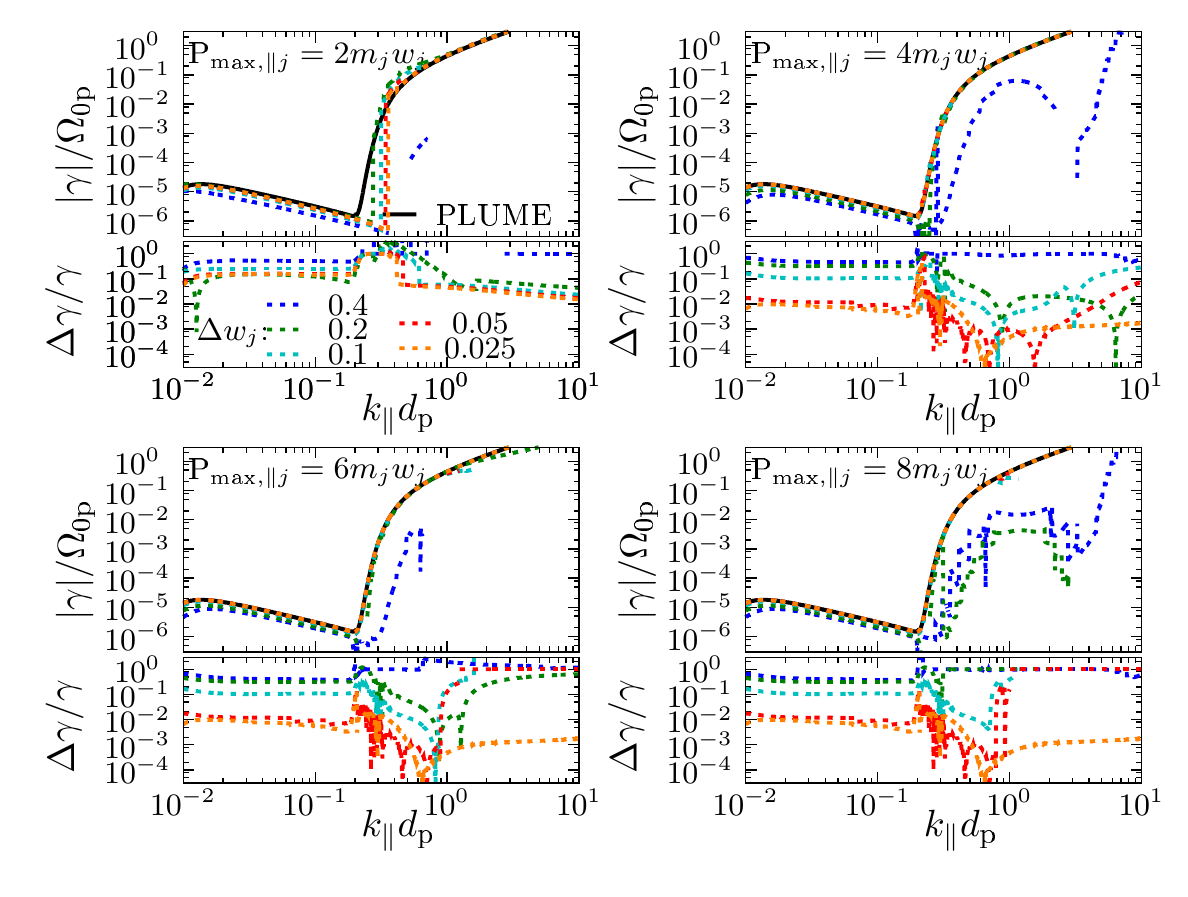}}
  \caption{Resolution study for the imaginary part of the frequency for the A/IC-wave solution in quasi-parallel propagation. We keep $k_{\perp}d_{\mathrm p}=10^{-3}$ constant and scan through $k_{\parallel}$. }
\label{fig_ALPS_convergence_a_par_gamma}
\end{figure}

In Figure~\ref{fig_ALPS_convergence_a_par_omega}, we show a resolution study for the A/IC wave in quasi-parallel propagation in an isotropic Maxwellian plasma. This figure complements our solutions shown in Figure~\ref{fig_par_example}. The four panels represent different values of $P_{\max,\parallel j}$. In each panel, the diagram at the top compares the real part of the frequency from five ALPS calculations with different $\Delta w_{j}$ to the Maxwellian solutions from PLUME. The diagram at the bottom compares the ratio between $\omega_{\mathrm r}$ from the five ALPS calculations and $\omega_{\mathrm r}$ from PLUME. In this parameter range, $P_{\max,\parallel\mathrm p}=8m_{\mathrm p}w_{\mathrm p}$ with a resolution finer than $\Delta w_j=0.1$ leads to a very good agreement with the PLUME solutions for $\omega_{\mathrm r}$. For wavenumbers below $1/d_{\mathrm p}$, a lower value of $P_{\max,\parallel\mathrm p}$ is sufficient. Figure~\ref{fig_ALPS_convergence_a_par_gamma} shows the same as Figure~\ref{fig_ALPS_convergence_a_par_omega}, but giving the imaginary part of the frequency instead of its real part. This figure confirms our finding regarding the optimal resolution. 

\begin{figure}
  \centerline{\includegraphics[trim={0.4cm 0.5cm 0.2cm 0.1cm},clip,width=\textwidth]{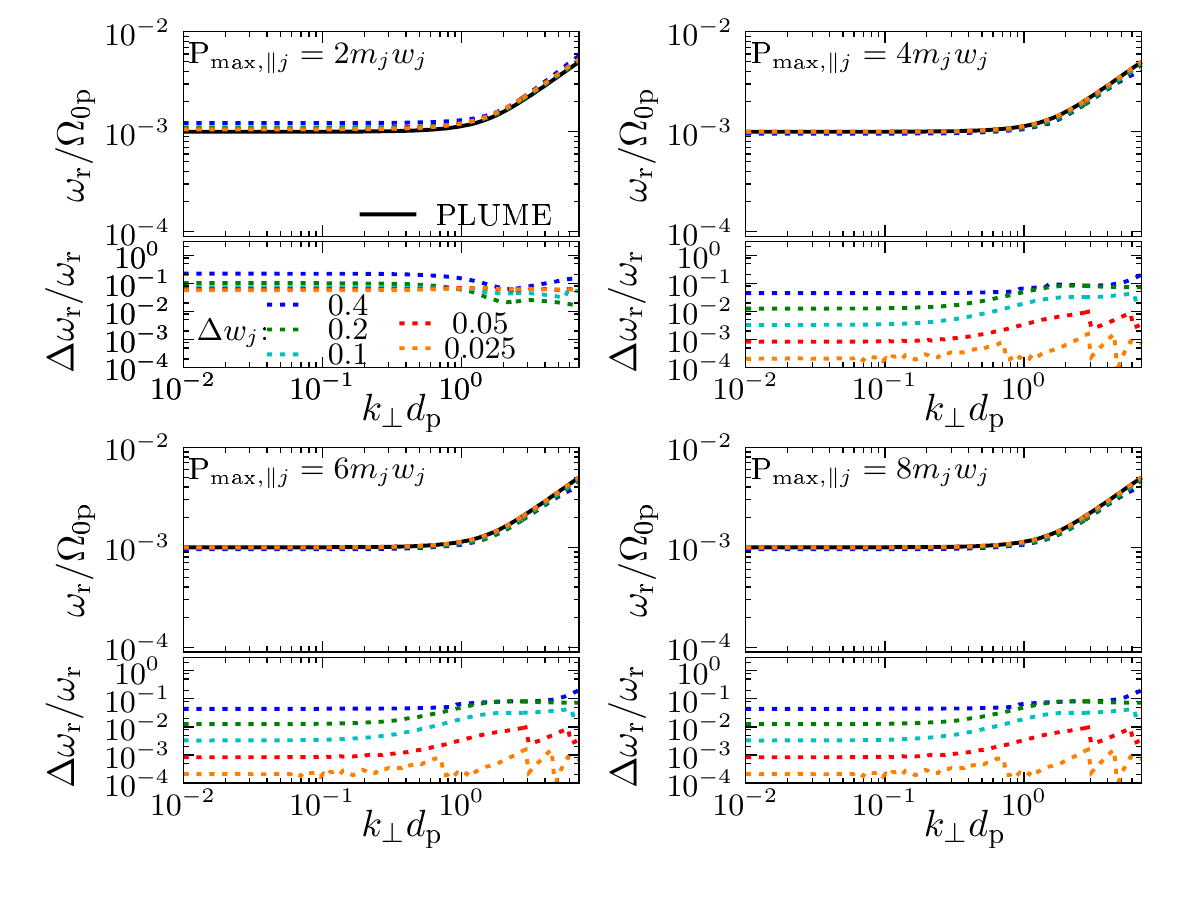}}
  \caption{Resolution study for the real part of the frequency for the A/IC-wave solution in quasi-perpendicular propagation. We keep $k_{\parallel}d_{\mathrm p}=10^{-3}$ constant and scan through $k_{\perp}$. }
\label{fig_ALPS_convergence_a_perp_omega}
\end{figure}

\begin{figure}
  \centerline{\includegraphics[trim={0.4cm 0.5cm 0.2cm 0.1cm},clip,width=\textwidth]{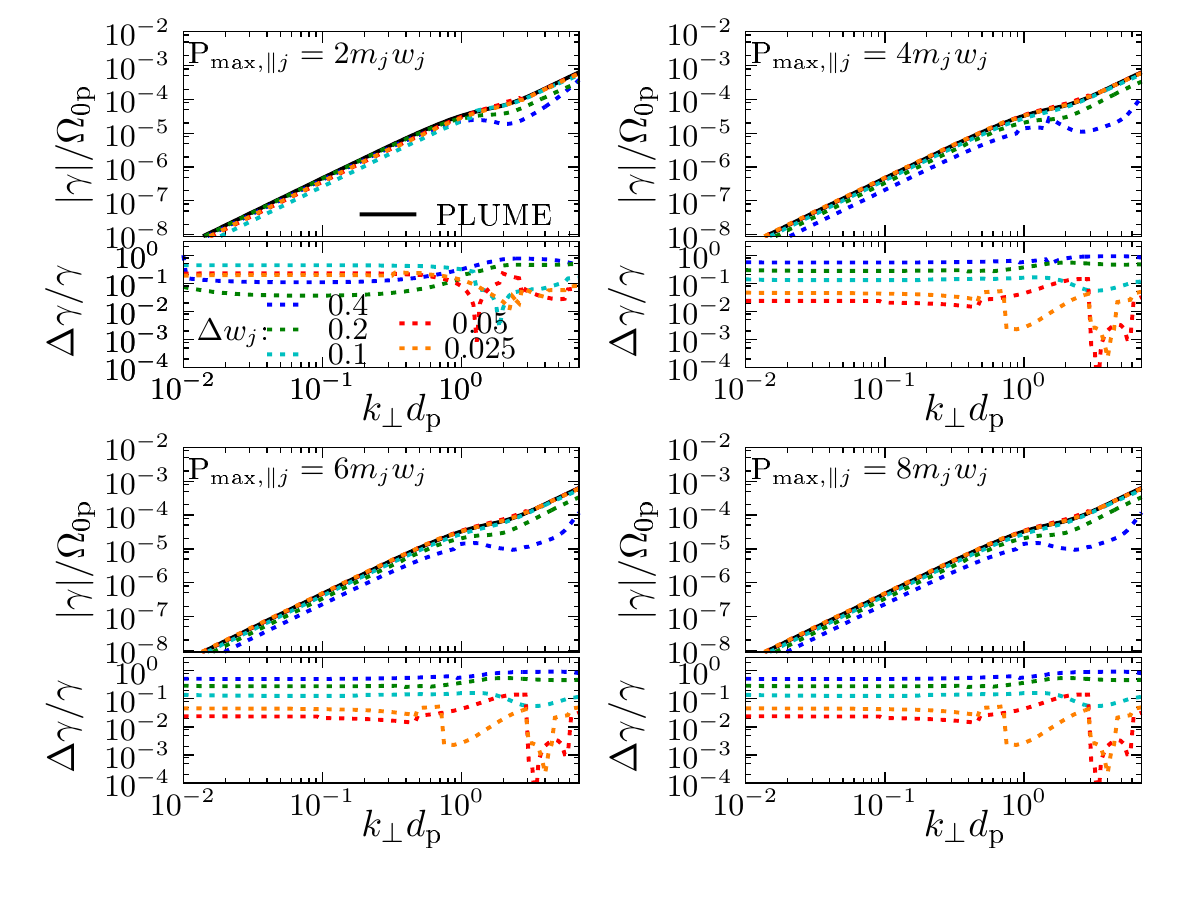}}
  \caption{Resolution study for the imaginary part of the frequency for the A/IC-wave solution in quasi-perpendicular propagation. We keep $k_{\parallel}d_{\mathrm p}=10^{-3}$ constant and scan through $k_{\perp}$. }
\label{fig_ALPS_convergence_a_perp_gamma}
\end{figure}

Figures~\ref{fig_ALPS_convergence_a_perp_omega} and \ref{fig_ALPS_convergence_a_perp_gamma} show the same as Figures~\ref{fig_ALPS_convergence_a_par_omega} and \ref{fig_ALPS_convergence_a_par_gamma}, but for quasi-perpendicular propagation instead of quasi-parallel propagation. The required resolution is lower in the quasi-perpendicular case than in the quasi-parallel case. The solutions with $P_{\max,\parallel\mathrm p}=4m_{\mathrm p}w_{\mathrm p}$ and $\Delta w_j\leq 0.1$ lead to a very good agreement between the ALPS and PLUME solutions.

\subsection{Anisotropic Bi-Maxwellian Distributions}\label{app_resolution_instabilities}

\begin{figure}
  \centerline{\includegraphics[trim={0.4cm 0.5cm 0.2cm 0.1cm},clip,width=\textwidth]{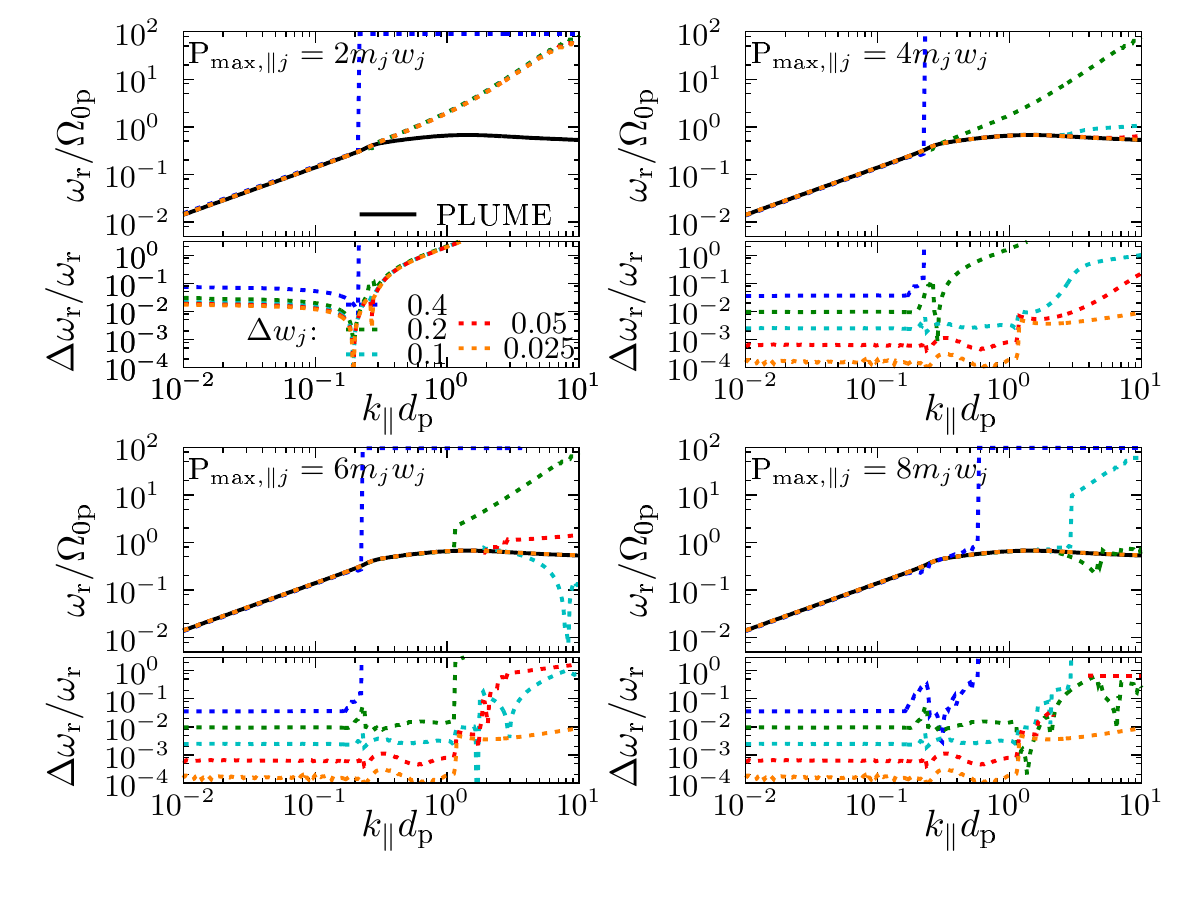}}
  \caption{Resolution study for the real part of the frequency for the A/IC-instability solution in quasi-parallel propagation. We use a bi-Maxwellian plasma with $T_{\perp\mathrm p}/T_{\parallel\mathrm p}=3$. We keep $k_{\perp}d_{\mathrm p}=10^{-3}$ constant and scan through $k_{\parallel}$. }
\label{fig_ALPS_convergence_pci_omega}
\end{figure}

\begin{figure}
  \centerline{\includegraphics[trim={0.4cm 0.5cm 0.2cm 0.1cm},clip,width=\textwidth]{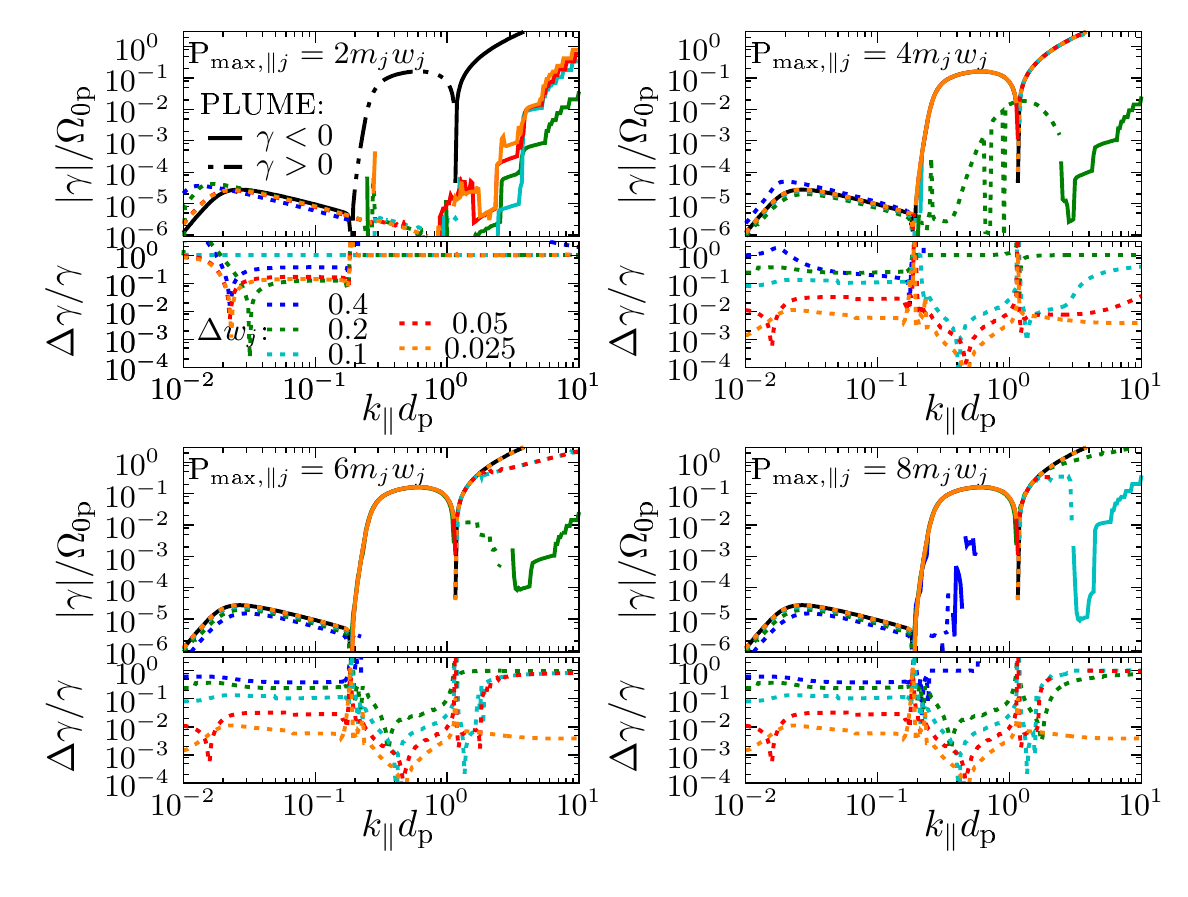}}
  \caption{Resolution study for the imaginary part of the frequency for the A/IC-instability solution in quasi-parallel propagation. We use a bi-Maxwellian plasma with $T_{\perp\mathrm p}/T_{\parallel\mathrm p}=3$. We keep $k_{\perp}d_{\mathrm p}=10^{-3}$ constant and scan through $k_{\parallel}$. }
\label{fig_ALPS_convergence_pci_gamma}
\end{figure}

In addition to our Maxwellian test, we study the dependence of the ALPS solutions on the resolutions for the bi-Maxwellian case with $T_{\perp \mathrm p}/T_{\parallel\mathrm p}=3$ as shown in Figure~\ref{fig_instable_example}. Figure~\ref{fig_ALPS_convergence_pci_omega} compares ALPS solutions for the A/IC instability in quasi-parallel propagation for different values of $P_{\max,\parallel\mathrm p}$ and $\Delta w_{j}$ with the solutions from PLUME for the real part of the frequency.  Figure~\ref{fig_ALPS_convergence_pci_gamma} compares ALPS and PLUME solutions for the imaginary part of the frequency. The solutions with $P_{\max,\parallel\mathrm p}=4m_{\mathrm p}w_{\parallel\mathrm p}$ and $\Delta w_j\leq 0.05$ lead to a good agreement between ALPS and PLUME in both $\omega_{\mathrm r}$ and $\gamma$.

\begin{figure}
  \centerline{\includegraphics[trim={0.3cm 0.1cm 0.2cm 0.1cm},clip,width=\textwidth]{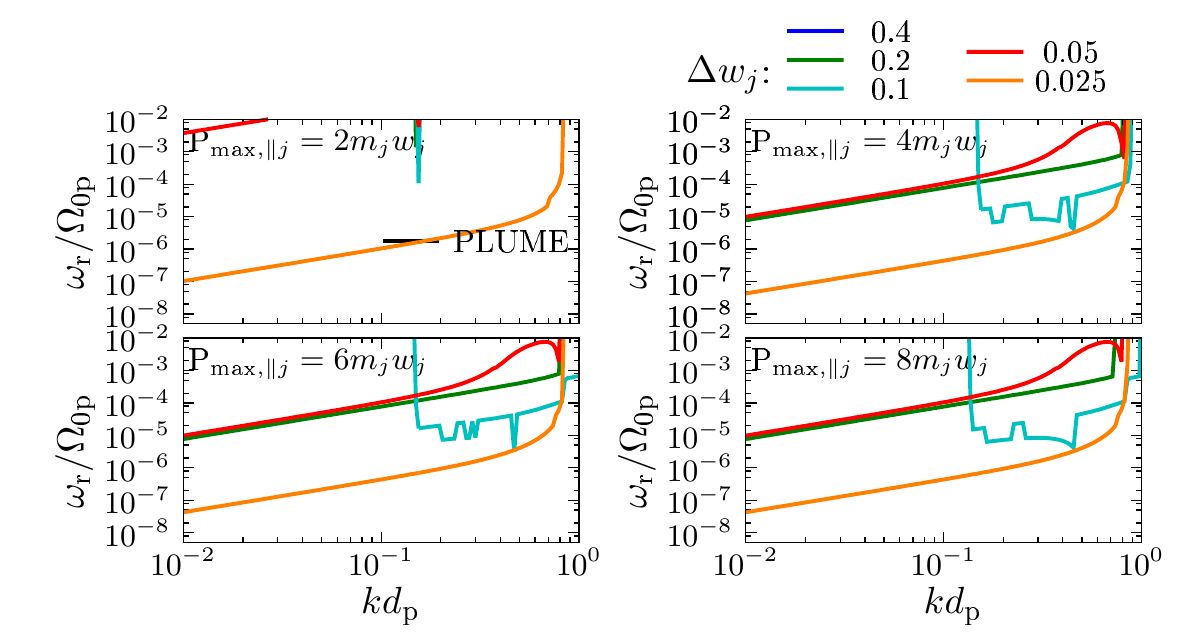}}
  \caption{Resolution study for the real part of the frequency for the mirror-mode-instability solution. We use a bi-Maxwellian plasma with $T_{\perp\mathrm p}/T_{\parallel\mathrm p}=3$. We keep $\theta=75^{\circ}$ constant and scan through $|\vec k|$. }
\label{fig_ALPS_convergence_mir_omega}
\end{figure}

\begin{figure}
  \centerline{\includegraphics[trim={0.4cm 0.3cm 0.2cm 0.2cm},clip,width=\textwidth]{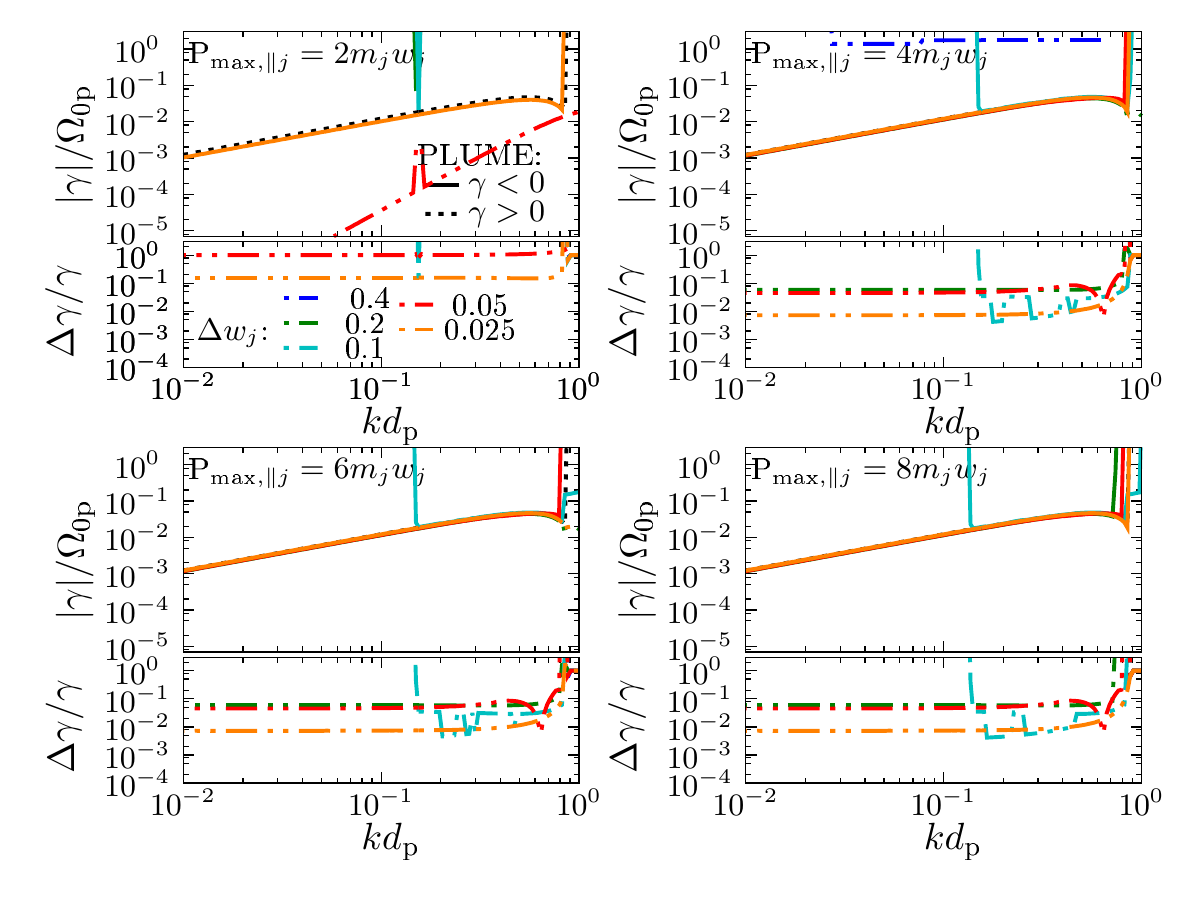}}
  \caption{Resolution study for the imaginary part of the frequency for the mirror-mode-instability solution. We use a bi-Maxwellian plasma with $T_{\perp\mathrm p}/T_{\parallel\mathrm p}=3$. We keep $\theta=75^{\circ}$ constant and scan through $|\vec k|$. }
\label{fig_ALPS_convergence_mir_gamma}
\end{figure}

In Figure~\ref{fig_ALPS_convergence_mir_omega}, we study the dependence of the solutions on the resolution for the mirror-mode instability with the same parameters as in Figure~\ref{fig_instable_example}. The correct solution of the mirror-mode instability has $\omega_{\mathrm r}=0$; however, the ALPS solutions have finite values $\omega_{\mathrm r}\neq 0$. The value of $\omega_{\mathrm r}$ decreases with increasing $\Delta w_j$. As \citet{Southwood:1993} point out, the mirror-mode instability is strongly influenced by particles with $p_{\parallel }\approx 0$. The error in frequency $\Delta \omega_{\mathrm r}$ is determined by the resolution of the momentum grid around $p_{\parallel}=0$, where $\Delta \omega_{\mathrm r}\sim k_{\parallel}w_j\, \Delta w_j$. Figure~\ref{fig_ALPS_convergence_mir_gamma} shows the comparison of the imaginary part of the mirror-mode solutions. Like in the case of the A/IC instability, a resolution with $P_{\max,\parallel\mathrm p}=4m_{\mathrm p}w_{\parallel\mathrm p}$ and $\Delta w_j\leq 0.05$ leads to a good agreement between ALPS and PLUME.

\section{Levenberg--Marquardt Fit}\label{sect_LMfit}

For the hybrid analytic continuation, ALPS fits the $f_0$ table with a combination of pre-described algebraic expressions as described in Section~\ref{hac}. 
We employ a Levenberg--Marquardt algorithm \citep{Levenberg:1944,Marquardt:1963} to fit the distribution functions in $p_{\parallel}$ with a superposition of an arbitrary number of Maxwellian distributions, $\kappa$-distributions,  and J\"uttner distributions. The user can freely choose the number of fits and their superposition. We evaluate different fit parameters for each given value of $p_{\perp}$.
We define the Maxwellian fitting function as
\begin{equation}
F_{\mathrm M}(\hat p_{\parallel})=u_{1}\exp\left[-y\hat p_{\perp}-u_{2}\left( \hat p_{\parallel}-u_{3}\right)^2\right],
\end{equation}
where $u_k$ are the fit parameters, $y$ is a constant user-defined parameter, and $\hat p_{\perp}$ and $\hat p_{\parallel}$ are the normalised perpendicular and parallel momenta. The parameter $y$ compensates the otherwise strong $p_{\perp}$-dependence of $u_1$, making the fit more reliable. It is constant for all $p_{\perp}$. We choose this expression rather than a fit in $p_{\perp}$ since it provides a greater flexibility in the $p_{\perp}$-domain compared to a two-dimensional fit in $p_{\perp}$ and $p_{\parallel}$. The best choice for $y$ is $\beta_{\perp j}m_{\mathrm p}/m_j$. The standard normalisation in ALPS uses $\hat p_{\perp}=p_{\perp}/m_{\mathrm p}v_{\mathrm A}$ and $\hat p_{\parallel}=p_{\parallel}/m_{\mathrm p}v_{\mathrm A}$. In cases with $\kappa$-distributed plasma components, we use 
\begin{equation}
F_{\kappa}(\hat p_{\parallel})=u_1\left[1+u_{2}\left(\hat p_{\parallel}-u_3\right)^2+y\hat p_{\perp}^2\right]^{u_4}.
\end{equation}
In cases with J\"uttner-distributed plasma components, we use
\begin{equation}
F_{\mathrm J}=u_1 \exp \left(-y \Gamma \right).
\end{equation}
In this case, the best choice for $y$ is $2c^2/w_j^2$.
These fitting relations are easily extendable by the user to cover more general functions as needed.

We denote the discretised $f_0$ table of species $j$ at constant $p_{\perp}$ as $\hat f_{i,j}(\hat p_{\parallel,i})$, the discrete steps in $\hat p_{\parallel}$ as $\hat p_{\parallel,i}$, the vector of all fit parameters as $\vec u$, and the sum of all fit functions as $F(\hat p_{\parallel})$. In the J\"uttner-distributed cases, the coordinates are replaced with $\Gamma$ and $\bar p_{\parallel}$ accordingly.
We define the residuals as $s_i\equiv \hat f_{i,j}-F(\hat p_{\parallel,i})$ and define $C\equiv \sum_is_i^2$.
We denote the Jacobian of $\hat{\vec f}_j$ with respect to $\vec u$ as $\mathsfbi J$. We use a superposition of analytical expressions for the Jacobian based on the given form of $F(\hat p_{\parallel})$.

The Levenberg--Marquardt algorithm uses an iterative step to update $\vec u$ of the form
\begin{equation}\label{LMproc}
\vec u_{\mathrm{new}}= \vec u + \left[\mathsfbi J^{\intercal}\mathsfbi J+\lambda\text{ diag}(\mathsfbi J^{\intercal}\mathsfbi J)\right]^{-1}\mathsfbi J^{\intercal}\vec s,
\end{equation}
where $\lambda$ is a user-defined scalar. For the matrix inversion in Equation~(\ref{LMproc}), we use the $\mathsfbi{LU}$-factorisation. Then we calculate the residuals $\vec s_{\mathrm{new}}$ based on $\vec u_{\mathrm{new}}$ and determine $C_{\mathrm{new}}=|\vec s_{\mathrm{new}}|^2$. If $C_{\mathrm{new}}\leq C$, we set $\vec u$ to $\vec u_{\mathrm{new}}$, reduce $\lambda$ by a constant factor $\lambda_{\mathrm f}$ (user-defined, standard value is 10), and repeat the procedure. If $C_{\mathrm{new}}>C$, we discard $u_{\mathrm{new}}$, increase $\lambda$ by the constant factor $\lambda_{\mathrm f}$, and repeat the procedure. In this way, we iteratively determine the fit parameters $\vec u$ until   the fit converges (i.e., $C\leq \epsilon$ with a user-defined $\epsilon$), or until the number of iterations reaches a user-defined maximum value. ALPS writes the fitted distribution into a separate output file so that a direct comparison with the original input distribution is possible.

\section{The Smoothed Thin-Plate Spline Interpolation}\label{interpolation}

Spacecraft or other plasma data are typically not available on a dense Cartesian grid like the grid required for an $f_0$ table in ALPS. Therefore, our code includes an interpolation algorithm that  fills gaps between data points. ALPS uses the same interpolation algorithm to create an equidistant grid in $(\Gamma,\bar p_{\parallel})$ space after the coordinate transformation in cases with relativistic poles.
We use a polyharmonic spline interpolation with the radial basis function of a thin-plate spline with smoothing \citep{powell94,donato02}. 
For each species, we begin with the ``coarse'' distribution function $\hat f_{\mathrm c,\mu}$ which is given by $n_{\mathrm c}$ data points (index $\mu=1\dots n_{\mathrm c}$) with the associated coarse momentum coordinates $\hat p_{\perp\mathrm c,\mu}$ and $\hat p_{\parallel\mathrm c,\mu}$. The set $(\hat f_{\mathrm c,\mu},\hat p_{\perp\mathrm c,\mu},\hat p_{\parallel\mathrm c,\mu})$ forms one data point. The coarse grid is typically not equally distributed in momentum space.

For each species, the ``fine'' grid of momentum coordinates is given by $\hat p_{\perp,i,k}$ and $\hat p_{\parallel,i,k}$ with $i=1\dots n_{\perp}$ and $k=1\dots n_{\parallel}$ (and correspondingly in the coordinates $\Gamma$ and $\bar p_{\parallel}$ for cases with relativistic poles). The fine grid corresponds to the actual $f_0$ table to be used as input in ALPS. The goal of our interpolation is to find the value of the distribution function $\hat f_{i,k}$ on all grid points $(i,k)$. We define the vectors $\vec w=(w_1,\dots,w_{n_{\mathrm c}})$, $\vec c=(c_1,c_2,c_3)$, $ \hat{\vec f}_{\mathrm c}=(\hat f_{\mathrm c,1},\dots,\hat f_{\mathrm c,n_{\mathrm c}})$, and $\vec 0=(0,0,0)$. We furthermore define the matrix 
\begin{equation}
K_{\mu,\nu}=\begin{cases}
   r^2\log (r)& \text{if } r\geq 1\\
    r\log(r^r)              & \text{if } r<1,
\end{cases}
\end{equation}
where $r\equiv \sqrt{(\hat p_{\perp\mathrm c,\mu}-\hat p_{\perp\mathrm c,\nu})^2+(\hat p_{\parallel\mathrm c,\mu}-\hat p_{\parallel\mathrm c,\nu})^2}$. We also define the $(n_{\mathrm c}\times 3)$ matrix $\mathsfbi P$. Its $\mu$th row is given by $(1,\hat p_{\perp\mathrm c,\mu},\hat p_{\parallel\mathrm c,\mu})$. The thin-plate spline interpolation requires to solve the nonhomogeneous linear system of equations
\begin{equation}\label{splinemat}
\begin{pmatrix}
\begin{array}{c|c}
\mathsfbi K +\alpha \mathsfbi 1& \mathsfbi P \\ \hline
\mathsfbi P^{\intercal} & 0
\end{array}
\end{pmatrix}
\begin{pmatrix}
\begin{array}{c}
\vec w \\ \hline
\vec c
\end{array}
\end{pmatrix}=
\begin{pmatrix}
\begin{array}{c}
\hat{\vec f}_{\mathrm c} \\ \hline
\vec 0
\end{array}
\end{pmatrix}
\end{equation}
for the vectors $\vec w$ and $\vec c$. $\alpha$ is a user-defined smoothing parameter ($\alpha=0$ forces the fine grid to run through all points of the coarse grid), and $\mathsfbi 1$ is the $(n_{\mathrm c}\times n_{\mathrm c})$ unit matrix. The interpolation is then given by
\begin{equation}
\hat f_{i,k}=c_1+c_2\hat p_{\perp,i,k}+c_3\hat p_{\parallel,i,k}+\sum\limits_{\mu=1}^{n_{\mathrm c}}w_{\ell} R_{i,k}^{\mu},
\end{equation}
where $R_{i,k}^{\mu}\equiv \sqrt{(\hat p_{\perp,i,k}-\hat p_{\perp\mathrm c,\mu})^2+(\hat p_{\parallel,i,k}-\hat p_{\parallel\mathrm c,\mu})^2}$.
The numerically expensive part of the interpolation is the solution of Equation~(\ref{splinemat}). Since $K_{11}=0$, a direct $\mathsfbi{LU}$ factorisation is not possible. Therefore, we apply a $\mathsfbi{LU}$-factorisation algorithm with partial pivoting through row permutations until $K_{11}\neq 0$.

\bibliographystyle{jpp}
\bibliography{ALPS}

\end{document}